\documentclass{article}

\usepackage{PRIMEarxiv}

\usepackage{subcaption}
\usepackage{graphicx}
\usepackage{multirow}
\usepackage{booktabs}
\usepackage{diagbox} 
\usepackage{longtable}
\usepackage{array}
\usepackage{rotating}
\usepackage{eqparbox}
\usepackage{makecell, caption, booktabs}
\usepackage{url}

\newcommand{\res}[2]{\small{$#1$}{\tiny{$\pm #2$}}}

\newcommand\feature[1]{{\small{\fontfamily{qhv}\selectfont #1}}}

\usepackage{pifont}
\newcommand{\cmark}{\ding{51}}%

\title{You Can't Hide Behind Your Headset:\\User Profiling in Augmented and Virtual Reality
}

\author{
  Pier Paolo Tricomi, Federica Nenna, Luca Pajola\thanks{First three authors contributed equally to this work.} \\
  University of Padova \\
  Padova, Italy\\
  \texttt{\{name.surname\}@studenti.unipd.it}
   \And
   Mauro Conti, Luciano Gamberini\\
  University of Padova \\
  Padova, Italy\\
  \texttt{\{surname\}@unipd.it} 
}

\begin{document}
\maketitle

\begin{abstract}
  Virtual and Augmented Reality (VR, AR) are increasingly gaining traction thanks to their technical advancement and the need for remote connections, recently accentuated by the pandemic.
Remote surgery, telerobotics, and virtual offices are only some examples of their successes. 
As users interact with VR/AR, they generate extensive behavioral data usually leveraged for measuring human behavior. However, little is known about how this data can be used for other purposes.  
\par
In this work, we demonstrate the feasibility of user profiling in two different use-cases of virtual technologies: AR everyday application ($N=34$) and VR robot teleoperation ($N=35$).
Specifically, we leverage machine learning to identify users and infer their individual attributes (i.e., age, gender). 
By monitoring users' head, controller, and eye movements, we investigate the ease of profiling on several tasks (e.g., walking, looking, typing) under different mental loads.
Our contribution gives significant insights into user profiling in virtual environments. 
\end{abstract}


\section{Introduction}

In recent years, the pandemic has increased the need of remote connections, and we have witnessed to a mass adoption of virtual technologies particularly for teamwork. This has opened new perspectives for different platforms that allow virtual interactions with others, and fostered the already ascending development of the Metaverse. The Metaverse has been recently defined as a "post-reality universe, a perceptual and persistent multiuser environment merging physical reality with digital virtuality" \cite{mystakidis2022metaverse}. While being designed around the human, which constitutes the physical reality of this interplay, the digital virtuality relies on immersive technologies that allow spatial and interactive features, namely AR and VR. Eventually, these devices became the core of the fourth wave of computing innovation \cite{kamenov2017immersive}.
\par
\par
Currently, there is an ongoing discussion on the potential protocols that will govern the Metaverse, with a particular focus on the controversial interplay between openness and privacy \cite{mystakidis2022metaverse}. The latest virtual devices allow tracking a large number of behavioral metrics, such as the headset's and controllers' position and rotation (which reflect the users' physical actions), all the interactions between the user and any virtual object present in the scene, and also eye movements. All these data can be source of personal information, and even the user's identity (e.g.,~\cite{miller2020personal} \cite{pfeuffer2019behavioural} \cite{rogers2015approach}). While being private, this information would help to restrict the use of the headset to specific individuals. For example, it would be possible to allow authentication only to those who have the rights, thus increasing the security of such technologies.

\subsection{Contributions.}
In this study, we assessed the feasibility of profiling and identifying users by leveraging behavioral data generated during an AR and a VR headset. 
We propose a general profiling framework that could be applied to different virtual devices (i.e., VR, AR), different applied fields (i.e., everyday use-case of a smart technology, and a work scenario), and different type of user's behaviors (i.e., walking, searching for landmarks, pointing, performing controller-based operations and physical actions). Second, we deeply study users' profiling at different levels (i.e., identification, age, gender), introducing - to the best of our knowledge - the novelty of profiling personal information of users (specifically, gender and age) in virtual contexts. Third, we additionally explored the impact of each sensor in users' profiling both with AR and VR, specifically assessing the relevance of position and rotation of the headsets and controllers hardware, and of the eye-tracking technology embedded in the VR device.
More importantly, we fill a gap in the literature on users' profiling in AR scenarios: while it is true that AR technology is still immature, it is also true that it is largely understudied compared to VR. We summarize our contributions as follows:

\begin{itemize}
    \item we propose a general profiling framework for AR and VR technologies;
    \item we study users' profiling with respect to identification, age and gender inference in virtual contexts, which is novel in the AR context;
    \item we conduct extensive studies to assess sensor' importance in our profiling tasks.
\end{itemize}

\par
\subsection{Organization.} 
In Section~\ref{sec.rel_work}, we provide background and review literature on users' profiling. Section~\ref{sec.method} presents the general profiling framework we adopted in our experiments. The dataset and experimental settings are shown in Section~\ref{sec.dataset} and Section~\ref{sec.experiment}, respectively. We report our results in Section~\ref{sec.results}, and conclude with a discussion in Section~\ref{sec.concl}. 

\section{Background \& Related Work}\label{sec.rel_work}
This section aim to describe the importance of security and privacy in virtual technologies such as AR and VR. 
Section~\ref{ssec.rel_work/applications} summarizes the application of virtual technologies in different fields, from the industry to the medicine. 
Section~\ref{ssec.rel_works/privacy} introduces the threats to AR and VR application with a cyber-security perspective.
Section~\ref{ssec.rel_works/sota} describes literature of user profiling in virtual technologies.

\subsection{AR / VR use-cases in daily and work scenarios}\label{ssec.rel_work/applications}
\subsubsection{Industry and remote work}
With the advent of Industry 4.0, the benefits of virtual devices have been repeatedly shown in many domains: in the design cycle of products and manufacturing systems~\cite{berni2020applications}, for programming machines~\cite{malik2020virtual}, in the teleoperation industry~\cite{linn2017virtual,xiao2020three} and 
also for training novices~\cite{prattico2021towards,roldan2019training}. 
In any of these applications, virtual technologies allow the operator to perform work tasks while being immersed in virtual environment that faithfully emulates the physical one. This is particularly important also for architecture, engineering and construction experts: virtual technologies in this sector are helpful for stakeholder engagement, design support and review, construction planning and monitoring, management, and training \cite{delgado2020research}.
\par
Taken together, the reasons why to employ virtual technologies in industry are numerous. This will potentially lead to large-scale adoption of VR and AR devices in industrial fields, opening new questions about how to ensure individuals' security on the work place. An effective algorithm for automatically identifying workers wearing a headset might help in this direction. For example, it would be possible to enabling authentication only for those who have the rights on the workplace (e.g., site manager). Further, assuming that older workers might prefer a different design of the virtual environment \cite{liu2020you}, an accurate user profiling might help customizing virtual features on the user's age.

\subsubsection{Education}
As AR and VR have the potential to bridge the limitations of 2D e-learning environments, online education is one of the fundamental pivots of Metaverse \cite{mystakidis2022metaverse}. Literature extensively examined the characteristics that lead to the successful integration of immersive virtual technologies in education, as well as its positive influences on learning outcomes. For example, 17 positive effects of VR were identified for education, such as improving skills, living more realistic experiences, enhancing the intrinsic motivation and the level of interest in learning \cite{chavez2018virtual}; however, these effects were subject-specific. Additionally, \cite{radianti2020systematic} reviewed VR application studies by focusing on immersive VR environments for higher education. They showed that, even though most of the literature reports that VR for education is still in its experimental stages, there is a strong general interest in the use of immersive VR particularly in engineering, medicine and computer science education, and that this technology is mature enough for teaching procedural, practical and declarative knowledge.
\par
In view of a large-scale adoption of VR/AR for education, a viable profiling or identification algorithm certainly comes in handy. For example, automatic authentication can be efficient when VR/AR technologies are adopted in numerous classes. Similarly, it would be helpful to automatically detect a student's age for adapting lessons and virtual contents.

\subsubsection{Gaming and entertainment}
Virtual technologies play an essential role in the gaming market too. While VR games were already well spread since the 1990s (e.g., Virtual Reality Gear~\cite{ansari2022implementing}), from 2018, AR also reached a large entertainment crowd with the popularization of Pokémon Go, Snapchat, Apple's ARKit, and Google.com's ARCore~\cite{vasista2022augmented}. This sector is expected to grow exponentially, as it also embraced entertainment areas that go far beyond gaming and arcade: film and music industry, live show sectors and sports are just a few examples \cite{abdelmaged2021implementation}. Particularly after the expensive loss caused by the pandemic in these sectors, the development of immersive virtual platforms can help support the cinema, music and live-show industry \cite{ansari2022implementing}. For instance, VR cinema was deployed for movies, theatre and art exhibitions~\cite{sharma2022product}, providing users with a 360 leisure experience. Last, the recent explosion of virtual influencers phenomenon~\cite{conti2022virtual} confirms the crucial role of virtual technologies in both entertainment and marketing levels.

\par
It is clear how the user's profiling/identification could be used for marketing strategies in this sector (e.g., delivering customized advertising). Further, particularly in gaming platforms, user identification might help detect banned individuals and prevent their access to virtual games.

\subsubsection{Medicine}
Virtual technologies have been proven to be reliable medical tools both for doctors and patients. For instance, as it allows for simulate surgeries, VR can be beneficial for medical education and training \cite{yeung2021virtual}. Interestingly, AR has the potential to superimpose salient clinical records or visual aids supporting a surgery over the patient's body \cite{birlo2022utility}. Research on virtual control systems for remote robotic surgery operations is also growing \cite{taylor2016medical}. Further, from the patient's point of view, VR can help improve cognitive abilities after a traumatic brain injury \cite{maggio2022virtual}, or it can help increase engagement in Parkinson's motor training via gamification \cite{van2019effectiveness}.
\par
Under this view, detecting whether a user is the chief of surgery rather than a student can help restrict the rights during a surgical operation involving AR/VR. Similarly, profiling patients using a virtual headset could allow training customization and automatic recordings of clinical improvements.

\subsubsection{AR as a smart wearable technology} 
The latest AR smart glasses are fully wearable devices with computational functions. They allow users to download applications from a mobile operating system and provide various functionalities by freeing the user's hands \cite{kim2021applications}. 
Notably, most AR glasses currently on the market do not offer an integrated experience of social networks and streaming content. 
For instance, Vuzix developed AR smart glasses specifically designed to be used with drones or for navigation in unknown areas. AR as an assistive navigation device has also been tested in applied research \cite{zhao2019designing}. 
However, Facebook has already partnered with Ray-Ban and launched their Ray-Ban stories, which have raised important questions about ethical and privacy issues \cite{iqbal2022adopting}. Even though they currently do not allow projecting holograms in the field of view, this is an essential hint for possible connections between AR technology and social networks.
\par
In the foreseeable future, the next generation of smart glasses will likely allow projecting e-mails and notifications from social networks on the user's field of view. In this perspective, accurate automatic identification of the user during everyday activities could help restrict the visualization of personal messages only to the owner of the glasses.

\subsection{Privacy in Emerging Technologies}\label{ssec.rel_works/privacy}
The increasing popularity of big data~\cite{bigdata} coupled with the rapid adoption of various ``smart'' devices has resulted in parallel increases in privacy concerns. In today's society, most people consider data collection incessant and believe that the risks outweigh any benefits~\cite{riskspr}. To prevent (or at least reduce) the exposure of personal data, current and emerging technologies should support privacy by default~\cite{ozturk2021privacy}, in accordance with recent legislation such as GDPR~\cite{gdpr}. Fortunately, researchers are actively focusing on studying and adding a security and privacy level in emerging technologies. For instance,
Di Pietro and Cresci~\cite{9750221} deeply discussed security and privacy issues arising in the metaverse, allowing a better understanding and a consequent improvement of the technology with respect to its users. Similarly, Nair et al.~\cite{nair2022going}, proposed a system to browse metaverse in incognito, protecting their privacy from companies, surveillance agencies, or data brokers. Researchers have also focused on incorporating privacy-preserving measures on daily usage systems, such as authentication~\cite{barni2010privacy}, and more recently, de-authentication techniques\cite{cardaioli2022privacy}.

Besides protecting users' data from unwanted usage or sharing, past literature shows how attackers can use \textit{public} data in unconventional ways to profile users or to infer \textit{private} users' data (e.g., gender, age, personality traits). For instance, Conti and Tricomi~\cite{conti2020pvp} studied user profiling in video games, showing how public gaming data can be exploited to track gamers for malicious activities, e.g., harassment or cyberbullying.    Kosinski et al.~\cite{kosinski2013private} leveraged Facebook data to infer users' gender, age, personality, or sexual orientation.     
Jurgens et al.~\cite{jurgens2015geolocation} predicted people's physical locations from their tweets, while Zhang et al.~\cite{zhang2020practical} leveraged Sharing Platforms' reviews to predict users' gender. The results of such studies highlight the high risks connected with data availability and point to the need for further research to protect users' privacy better.

\subsection{Users Profiling in AR and VR applications}\label{ssec.rel_works/sota}
Privacy risks in AR and VR technologies is not deeply discussed in the current literature. 
Roger et al.~\cite{rogers2015approach} investigated the task of user identification, i.e., identifying a given user among a group of known people. This study is conducted in an AR environment through Google Glasses among 20 participants. Behavioral features include head movement (i.e., accelerometer and gyroscope) and eye blinking patterns.
The best performing model - a Random Forest - achieved 94\% of accuracy in the task.
Li et al.~\cite{li2016whose} proposed Headbanger, an authentication system for wearable devices. The authentication task differs from the identification one since in the first, users can be unknown, while in the latter, the algorithm aims to identify a user in a group of given users.
This study is conducted in an AR environment through Google Glasses among 95 participants. 
The proposed system relies on motion sensors (mainly the headset accelerometer), and the system authenticates users by leveraging three distance metrics, such as Cosine Distance, Correlation distance, and dynamic-time warping distance.  
Headbanger achieves 95\% of accuracy in the task.
Mustafa et al.~\cite{mustafa2018unsure} proposed an authentication system for VR, highlighting the importance of such a security mechanism, especially when a user is completely immersed in the virtual environment, which can lead to the dangerous \textit{lunch time attack}~\cite{eberz2015preventing}.\footnote{A \textit{lunch time attack} occurs when the victim walks away from the logged-in device, and thus an attacker can utilize such systems with the victim's privilege~\cite{conti2020auth}.}
This study was conducted through a Google Cardboard VR with a Samsung Galaxy S5 mounted and involved over 23 participants. 
Behavioral features involve sensors like the headset's accelerometer and gyroscope, from which the authors extracted features such as summary statistics (e.g., mean, variance) and frequency domain features (e.g., energy).
The best performing model - a Logistic Regression - achieve 93\% of accuracy in the task.
Pfeuffer et al.~\cite{pfeuffer2019behavioural} studied the problem of user identification in VR. 
The experiment is conducted with HTC Vive, involving 22 participants. 
The authors consider a broad spectrum of features that capture head, hand and eye motions.
The best performing model - a Random Forest - achieves up to 40\% of accuracy in the task.
Miller et al.~\cite{miller2020personal} further deeply explore the identification task in VR, as similarly done by Pfeuffer et al.~\cite{pfeuffer2019behavioural}.
The experiment is conducted with HTC Vive, involving 511 participants. 
Behavioral features include summary statistics (e.g., maximum, minimum, average, std), position and rotation of headset and controllers (both right and left hand).
The best performing model - a Random Forest - achieves up to 95\% of accuracy in the task.
\par
The reader can notice that existing works mainly focus on tasks inferring mainly the person with authentication or identification tasks, while there is a lack of understanding of wheter behavioral data can be leveraged to infer other users' private information such as age and gender. 
Similarly, prior works consider only AR or VR technologies for their experiments.
This is a limitation, since the level of virtual immersion allowed by the two technological devices is substantially different \cite{milgram1995augmented}, and this significantly affects the user's behaviors.
Our paper thus aims to fill the current literature gap by considering different privacy inference tasks (i.e., age, gender, identification) explored in both AR and VR environments. 
\begin{table}[ht]
    \centering
    \caption{State of the art overview.}
    \label{tab:sota}
    \footnotesize
    \resizebox{\columnwidth}{!}{%
    \begin{tabular}{cc|cc|ccccc} \toprule
         & & \multicolumn{2}{c|}{\textit{Technology}} & \multicolumn{4}{c}{\textit{Privacy-level}}\\
         \textbf{Reference}& \textbf{\#Participants}& \textbf{AR} & \textbf{VR} & \textbf{Age} & \textbf{Authentication} & \textbf{Gender} & \textbf{Identification} \\ \midrule
         Roger et al.~\cite{rogers2015approach}& 20 & Google Glass & & & & & \cmark\\
         Li et al.~\cite{li2016whose}& 95 & Google Glass & & & \cmark & & \\
         Mustafa et al.~\cite{mustafa2018unsure}& 23 & & Google Cardboard VR & & \cmark & & \\
         Pfeuffer et al.~\cite{pfeuffer2019behavioural}& 22 & & HTC Vive & & & & \cmark\\
         Miller et al.~\cite{miller2020personal}& 511 & & HTC Vive & & & & \cmark\\\midrule
         \textit{Our}& 34 (AR) and 35 (VR) & Microsoft HoloLens  & HTC VIVE Pro & \cmark & & \cmark & \cmark\\ \bottomrule
    \end{tabular}
    }
\end{table}

\section{Methodology}\label{sec.method}
This section describes the methodology we propose to execute inference task with virtual technologies.
Section~\ref{ssec.method/overview} motivates the reasons of our investigation.
Section~\ref{ssec.method/framework} presents our proposed framework. 

\subsection{Scope of the work}\label{ssec.method/overview}
Augmented Reality (AR) and Virtual Reality (VR) devices contain several sensors (e.g., accelerometer, gyroscope, eye tracking) essential to interact with virtual environments. 
Sensors' data describing human behaviour can be used to build biometric applications, opening several opportunities to enhance and tailor users' experience. However, such data might pose risks for users' privacy and security. 
In this study, we aim to understand whether it is possible to profile users by leveraging their interaction with AR and VR applications. In particular, we conduct our study by considering two categories of profiling:
\begin{enumerate}
    \item \textit{User identification}, where we aim to identify a given user within a known population;
    \item \textit{Private information inference}, where we aim to infer users' gender and age.
\end{enumerate}
We thus propose a general framework to accomplish both tasks, which can be extended to infer additional users' information. 

\subsection{Inference Framework}\label{ssec.method/framework}
\subsubsection{Overview}
\begin{figure}[!ht]
    \centering
    \includegraphics[width=.9\textwidth]{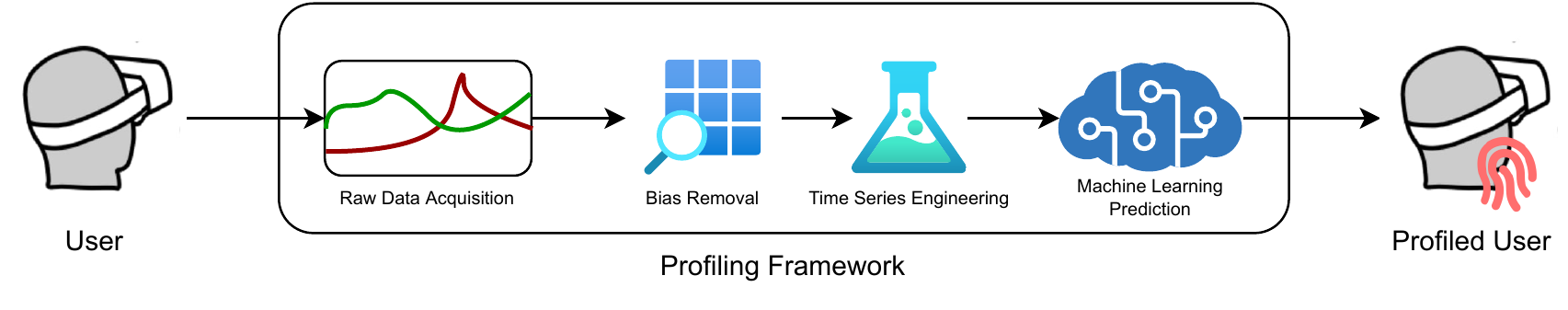}
    \caption{Overview of the proposed framework for user profiling in Augmented and Virtual Reality.}
    \label{fig:pipeline}
\end{figure}

Our goal is to define a generic pipeline that can be adapted and applied on any virtual technology (e.g., AR, VR) context to infer users' private information. 
As shown in Figure~\ref{fig:pipeline}, the pipeline consists of four steps, starting from the \textit{user} from whom we record the behaviors, to his/her actual profiling:
\begin{enumerate}
    \item \textit{Raw Data Acquisition}. In this phase, users' behavioral data are acquired. Virtual technologies' devices continuously generate data from users' interactions with the virtual environment (i.e., time series). 
    From these data, we can describe users' behavior. The amount and type of information depends on the virtual technology and its devices. For instance, data might come from users' input (e.g., pressing joystick's buttons) and users' movements.
    \item \textit{Bias Removal}. This phase aims to remove potential biases from time series that might lead to train erroneous machine learning models.  
    \item \textit{Time Series Engineering}. This phase aims to extract insightful information from the time series. 
    \item \textit{Machine Learning Prediction}. This phase aims to infer users' private information from the data elaborated in the previous phase by leveraging machine learning algorithms. 
\end{enumerate}

\paragraph{Raw Data Acquisition}
Users interact with AR and VR applications through devices such as headsets and joysticks. 
Such devices embeds several functional sensors to offer users an immersive experience. 
For example, users move and explore the virtual environment through sensors like accelerometer and gyroscope embedded in the headset.
Thus, by combining information retrievable by each sensor $s^i$ of the equipment, we can trace users activity $a$ at a given time $t$:
\begin{equation}\label{eq.timestamp}
    \vec{a_t} = [s^0_t, s^1_t, ..., s^n_t],
\end{equation}
where the subscript denotes the timestamp, and the superscript the sensor involved. 
We call this process \textit{acquisition phase}. 
Acquisition phase can be repeated over time, resulting in a user temporal behavioural description. Thus, by acquiring data in $\Delta t = t - t_0$, we obtain a behavioral time series, described as follows: 
\begin{equation}\label{eq.behavior}
    \vec{\mathbf{B}}_{\Delta t} = [\vec{a}_{t_0}, \vec{a}_{t_1}, ..., \vec{a}_{t-1},\vec{a}_t].
\end{equation}
$\vec{\mathbf{B}}_{\Delta t}$ represents an atomic sample of a user action (or task) of duration $ \Delta t$ that we will use in the next phases to infer their private information.

\paragraph{Bias Removal}
The acquisition phase might lead to enormous quantity of raw data. 
Such data might not only describe users behaviour, but also environmental information strongly correlated to experimental sessions. 
For example, using the raw headset height to identify users might be erroneous since such information might not be persistent over time (e.g., different shoes, different body position)~\cite{miller2020personal}. 
The problem of \textit{spurious correlations} in cybersecurity applications is well known~\cite{arp2022and}.
We thus need to be extra careful in understanding if sensors might lead to erroneous and inconsistent machine learning performance. 
The process of bias removal depends on the sensors' nautre and require an ad-hoc analysis. 
We explain in details our implementation in Section~\ref{ssec.experiment/implementation}.
The de-biasing phase results in a new vector of de-biased actions:
\begin{equation}\label{eq.debias}
    \vec{\mathbf{B}_{\Delta t}} = [\vec{d}_{t_0}, \vec{d}_{t_1}, ..., \vec{d}_{t - 1} ,\vec{d}_t],
\end{equation}
where $d_{t_i}$ is the de-biased version of the feature $a_{t_i}$.

\paragraph{Time Series Engineering}
Raw temporal data should be properly elaborated to extract meaningful information. 
Moreover, given the huge amount of data, such sequences should be aggregated (i.e., compressed) to limit the computational cost of their analyses.
The aggregation strategy can consider the whole sequence of a specific features, or just subpart of it.  
For example, given a sensor $s^i_{\Delta t}$ and its de-biased values over the time $d^i_{\Delta t} = [d_{t_0}^i, d_{t_1}^i, ..., d_{t - 1}^i ,d_t^i]$, the aggregation of a whole sequence results in a unique number $x^i$, while the partial aggregation (e.g., a transformation every $q$ times step) in a vector of numbers $[x^i_0, x^i_1, ..., x^i_m]$, where $m= t/q$. Note that the subscript does not denote anymore the temporal axis.  
Popular features derived from the aggregation phase are the mean, standard deviation, min, max~\cite{miller2020personal}. At the end of the process, we obtain, for each participant action or task, an aggregated datapoint that will be used by the machine learning models.


\paragraph{Machine Learning}
The last phase of the pipeline involves machine learning approaches like Logistic Regression (LR), Decision Tree (DT) and Random Forest (RF). 
Training a well-performing model requires validation strategies that consider the nature of the inference. 
For instance, if the aim is to identify a user within a known population, the training, validation and testing splits should contain samples of the population.
However, to avoid trials (or sessions) bias, the three split should consider samples belonging to different trials of collection. 
On the opposite, when inferring information like age and gender, the three splits should contain different set of users. 
Regarding the type of machine learning algorithm, we suggest the utilization of \textit{inherently interpretable} models (e.g, LR, DT) to better understand models decisions while inferring.  
Moreover, interpretable models allows a transparent debugging phase to identify the presence of spurious features~\cite{nadeem2022sok}.
Finally, given the unbalance nature of the problem (i.e, not all the classes are distributed equally), we suggest using performance metrics like F1-score with macro average.

\section{Dataset overview}\label{sec.dataset}
For the present investigation, we chose two use-case scenarios of virtual technologies, one involving AR and one VR. For both of them, we asked permission to the authors \cite{nenna2021augmented}, \cite{nenna2022virtualization} and \cite{nenna2022influence} for sharing their data with our team and conduct the present study. The first dataset, described in Seciton~\ref{ssec.dataset/ar}, thus comes from a study on the multitasking effects when using AR while walking outdoor \cite{nenna2021augmented}. Specifically, the authors took an experimental paradigm typically used in behavioral and cognitive research outside the lab, in a real dynamic scenario, and measured dual-task walking effects in young users responding to augmented stimuli during navigation. The second dataset, described in Section~\ref{ssec.dataset/vr}, instead comes from a use case scenario introducing VR into robotics and manufacturing industry \cite{nenna2022influence} \cite{nenna2022virtualization}. In this context, the authors tested users guiding an industrial robotic arm via different control systems in VR. Even though the allowed behaviors were kept as simple as possible to ensure experimental control, both scenarios give an important glimpse into practical applications of virtual technologies in the field. Furthermore, in both cases, the dual-task methodology was deployed for testing users' behavior under different levels of workloads. This traditional paradigm is extensively used in human factors and applied research and represents an ecologically valid but still control method for imposing mental strain on a user \cite{nenna2021augmented} \cite{nenna2022virtualization}.

\subsection{AR experiment}\label{ssec.dataset/ar}
\par
\begin{figure}[h!]
    \centering
    \includegraphics[width = 0.5\textwidth]{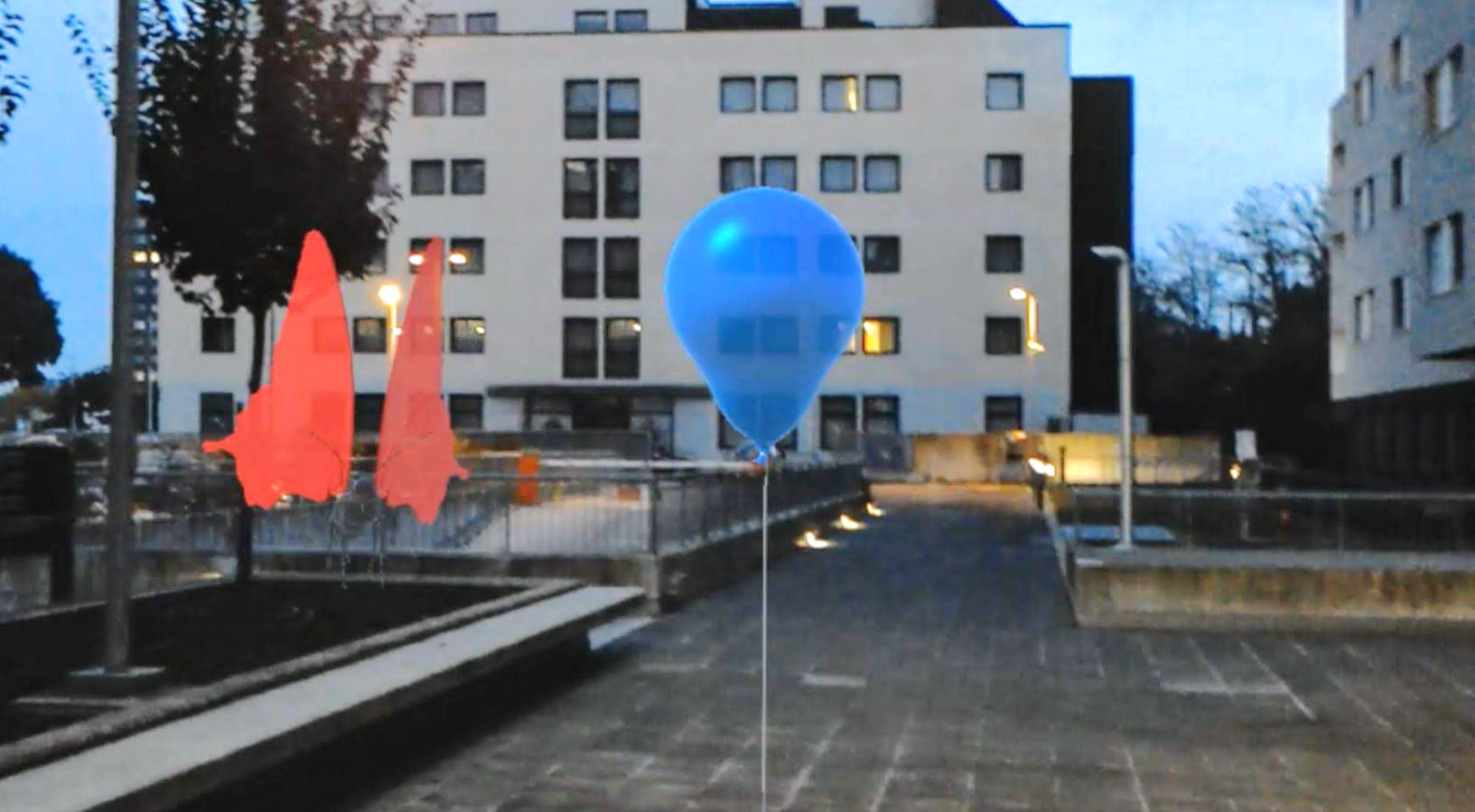}
    \caption{AR Environment.}
    \label{fig:AR_env}
\end{figure}

The AR experiment investigated multitasking effects in participants using AR while walking outdoor \cite{nenna2021augmented}. For this case study, 45 young adults wore the Microsoft HoloLens 1st generation smart glasses (OS Windows 10, CPU Intel 32-bit 1GHz, memory 2GB RAM and 1GB HPU RAM, 2.3 megapixel widescreen head-mounted display, field of view 30 × 17, mass 579g) and performed: i) a visual task, in which they discriminated between different augmented targets presented in their peripheral view, ii) a navigation task, in which they reached a series of augmented landmarks via physical walking outdoor, and iii) the combination of these tasks, which they called dual-task. The virtual environment, shown in figure~\ref{fig:AR_env}, was programmed in Unity (2017.4.18f1) and participants interacted with the augmented targets both via a wireless Xbox One controller and via physical collision with the virtual objects (e.g., walking through an augmented target). 
\par
Each participant performed 80 trials of the visual task, 50 trials of the navigation task and 50 trials of the dual-task. 
Specifically, for each trial of the visual task, a green or red object appeared lateralized on the left or on the right side of the visual field for 300ms; hence, the participant was asked to press a specific button on the joystick based on the color of the target and independently from the hemifield where it appeared. 
Differently, in the navigation task, a series of augmented landmarks appeared one after the other at -90°, 0° or 90° with respect to the participant's position and at a distance of 3m from each other. Participants were thus instructed to first inspect the surrounding to find the landmark, and then walk through it. 
In the dual-task, finally, participants walked through the series of landmarks while concurrently responding to the lateralized augmented stimuli. These tasks were specifically designed for measuring the effects of multitasking outside the lab. Therefore, they offer good insights into the potential impact of AR during outdoor walking.
\par
The dataset is composed of 21 females (age mean = 24.28, SD = 2.22) and 24 males (age mean = 24, SD = 2.62) and comprises the following continuous measures: position (in meters) of the AR headset in the three axes (x, y, z), and rotation of the AR headset in Euler angles. Furthermore, time stamps of any button press on the joystick and any collision with virtual objects presented in the scene were also registered, even if they were not considered for the present work. It is to notice that, since the datasets of the first 11 participants did not include data on the headset position, we ran our investigation on 34 participants out of 45.

\subsection{VR experiment}\label{ssec.dataset/vr}
The VR experiment deployed a virtual reproduction of an industrial robotic arm (Universal Robot UR5) developed in Unity (version 2020.2.1f1) \cite{nenna2022virtualization} \cite{nenna2022influence}. The virtual environment was designed to test performance and eye parameters of users during a simulated teleoperation task. All participants wore an HTC VIVE Pro Eye VR device (resolution 1440x1600 pixels per eye, refresh rate 90Hz, field of view 110°, weight 555g) and were provided with both VR controllers.
\par
\begin{figure}[h!]
    \centering
    \includegraphics[width = 0.5\textwidth]{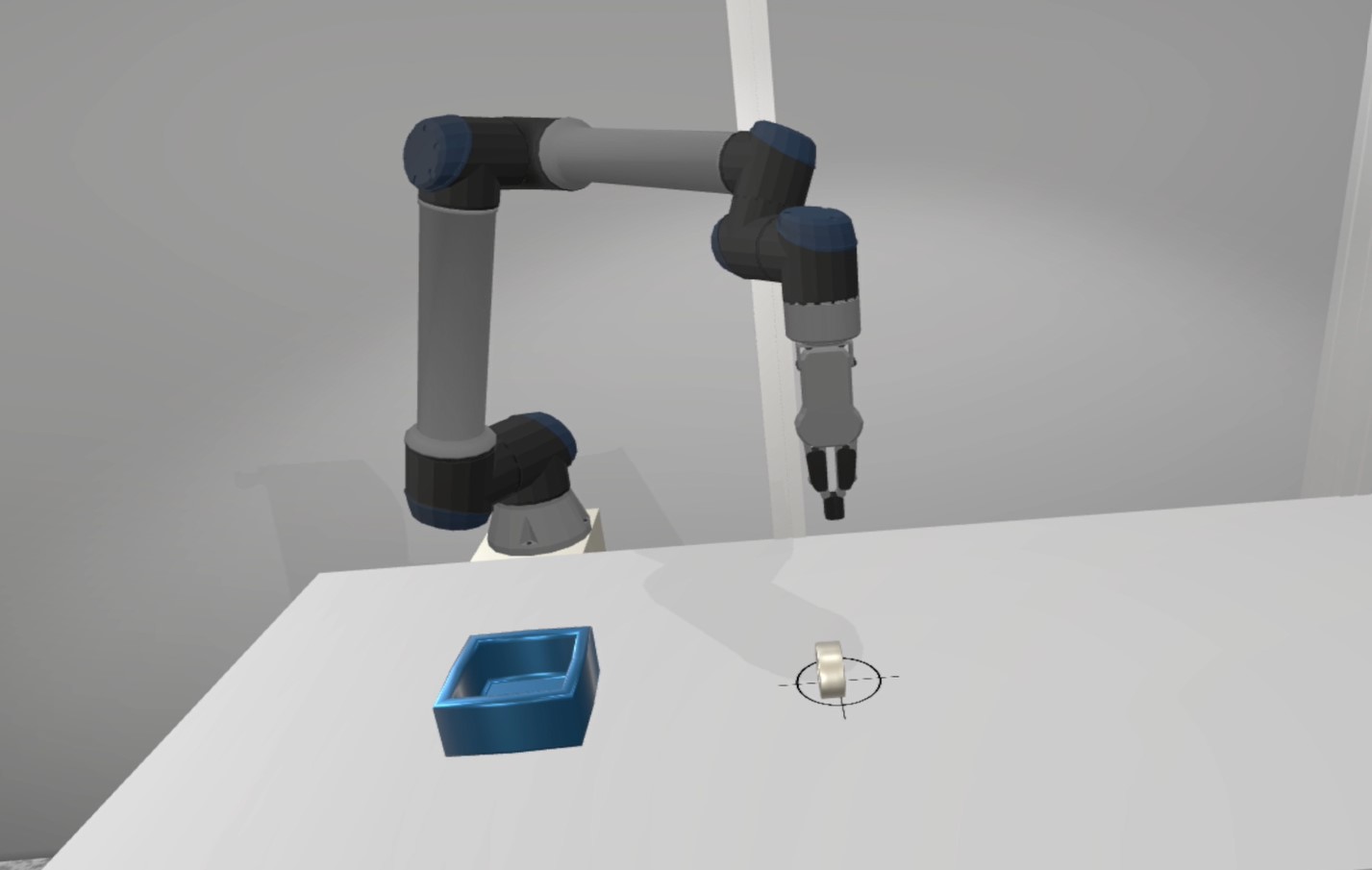}
    \caption{VR Environment}
    \label{fig:VR_env}
\end{figure}

The dataset included 21 young adults (10 females, 11 males) and 14 participants who reported being more than 50 years old (8 females, 6 males). Therefore, overall, 18 females (age mean = 39.33, SD = 14.21) and 17 males (age mean = 37.75, SD = 16.32) participated at the experiment. All participants in VR guided the robotic arm shown in figure~\ref{fig:VR_env} through a pick-and-place via two different control systems (controller buttons and physical actions) and under two levels of workload (single-task and dual-task). For the pick-and-place task, they had to pick a bolt from the workstation and place it into a box. When using the controller buttons system, they performed the task by only using the pad buttons on the VR controllers. With the physical actions system, instead, they still used the VR controllers, but they were allowed to physically approach the robot with their hand, grasp it and then move it over the worktable by physically moving their arm. Furthermore, in contrast with the single-task condition, in the dual-task, participants operated the pick-and-place task while also performing simple arithmetic sums. A series of numbers ranging between 1 and 10 were randomly presented on a virtual screen in front of the participant for the whole duration of the pick-and-place actions. 2.5s elapsed between each number presentation, with a random jitter of 0.3s. After the place action, participants reported the result of the arithmetic operation by pointing a virtual keyboard through the controller and then moved to the next trial. In each condition, the young participants performed 40 trials, while the old participants performed 20 trials.
\par
 The following continuous measures were registered: position (in meters) in the three axes (x, y, z) and rotation in Euler angles of both the VR headset and its controllers. As the VR device employed for these investigations is additionally provided with an integrated eye-tracker, the following continuous eye parameters were also recorded: pupil size (in millimeters) and eye openness (expressed from 0 to 1). Finally, time stamps of any button press on the controllers and any collision with virtual objects in the scene were registered too, but they were not used for the present investigation.

\section{Experimental setting}\label{sec.experiment}
This section describes our experimental settings. 
In particular, starting from the AR and VR datasets previously described in Section~\ref{sec.dataset}, we define inferring experiments based on the task-level (Section~\ref{ssec.experiment/task}) and action-level (Section~\ref{ssec.experiment/action}).
Section~\ref{ssec.experiment/implementation} describes the methodology we follow in our experiments (i.e., de-biasing, feature extraction, model selection).

\subsection{Task Identified from the AR and VR Experiments: Task-level}\label{ssec.experiment/task}
For the present investigation, we isolated specific macro tasks on which we performed users' profiling/identification. 

\subsubsection{Augmented Reality} 
Specifically, in AR we considered the same navigation task as identified by the authors \cite{nenna2021augmented}, and then we called mental task what the authors called visual discrimination task. In the latter, participants were discriminating between different colored and lateralized augmented objects while standing still. As this task was mentally demanding, we consider it as a mental task. Differently, in the navigation task, participants were looking for augmented targets in their surroundings and then walked through it. The navigation task was performed both as a single-task (low workload) and concurrently with the mental task (high workload). To recap, in AR environment, we identified the following tasks:
\begin{itemize}
    \item Mental Task (MT);
    \item Navigation Task - Low workload (NT-Low);
    \item Navigation Task - High workload (NT-High).
\end{itemize}

\subsubsection{Virtual Reality}
In VR, instead, we followed the same categorization used by the authors \cite{nenna2022influence}. Therefore, we considered two different pick-and-place tasks according to the type of interactions allowed between the user and the virtual robot: controller buttons and physical actions. The CB-based task corresponds to the pick-and-place performed via controller buttons, while the PA-based task includes the same pick-and-place executed via physical actions. Both tasks were executed under low and high workload: compared to the low workload, in the high workload condition participants executed the pick-and-place task simultaneously with the arithmetic task. Overall, the following tasks were identified from the VR scenario:
\begin{itemize}
    \item Controller-based Task - Low workload (CT-Low);
    \item Controller-based Task - High workload (CT-High);
    \item Action-based Task - Low workload (AT-Low);
    \item Action-based Task - High workload (AT-High).
\end{itemize}

\subsection{Actions Identified from the AR and VR Tasks: Action-level}\label{ssec.experiment/action}
Further, from each of the tasks discussed above, we identified a series of different actions both from the AR and VR experiments. The analysis performed on these actions is at a micro-level, and it is based on the type of interactions performed and on the range of motion involved.
\par
\subsubsection{Augmented Reality}
Specifically, from the tasks performed in AR, we extracted the following operations: button interaction, search and walk. 
In the button interaction, we included the task sections in which participants were standing still while discriminating between the lateralized colored targets. Specifically, they pressed specific buttons on the joystick according to the hemifield where the virtual object was displayed.
In the search operation, participants were engaged in the visual inspection of the surroundings for finding a virtual landmark; this operation was performed while participants were standing still and just rotated their head for inspecting the surrounding.
Finally, in the walking operation, participants were physically walking to the identified virtual landmark. 
Both the search and walk operations were performed as single-task and concurrently with the secondary mental task (namely, the visual discrimination task). As argued by the authors \cite{nenna2021augmented}, participants perceived lower workload when performing only the navigation task rather than performing the same task concurrently with the mental task. In other words, the secondary mental task put a strain on the users' mental resources. Therefore, we here refer to the dual-task as the high workload condition, while the single-task is considered as a low workload condition. Table~\ref{tab:AR_actions} represents the actions isolated in the AR environment.

\begin{table}[ht!]
\caption{Augmented Reality actions organized per type of operation and workload level.}
\label{tab:AR_actions}
\centering
\begin{tabular}{c|  >{\centering\arraybackslash}m{2cm}  >{\centering\arraybackslash}m{2cm}  >{\centering\arraybackslash}m{2cm}} \toprule     \multirow{2}{*}{\diagbox{\textit{Workload}}{ \textit{Operation}} } & \textbf{\textit{Button}} & \multirow{2}{*}{\textbf{\textit{Search}}} & \multirow{2}{*}{\textbf{\textit{Walk}}}   \\ 
    & \textbf{\textit{Interaction}} & &\\
    \toprule
  \textit{\textbf{Low}}
&  --  & \includegraphics[width=2cm]{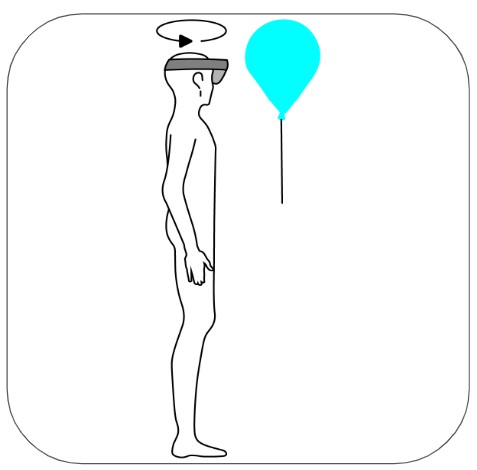} & \includegraphics[width=2cm]{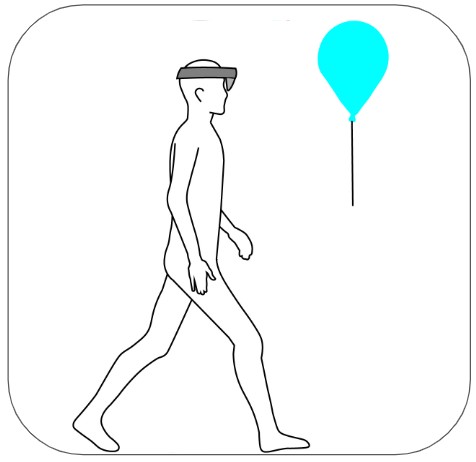}     \\
\textit{\textbf{High}} & \includegraphics[width=2cm]{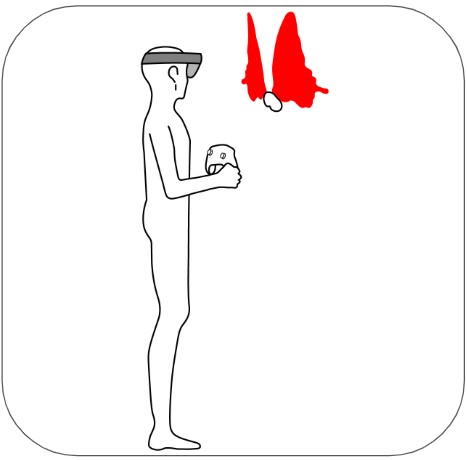}  &  \includegraphics[width=2cm]{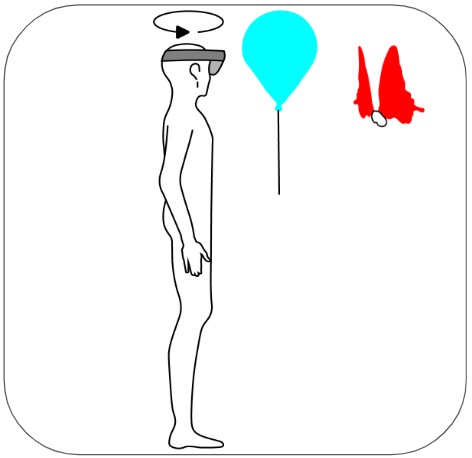}   &    \includegraphics[width=2cm]{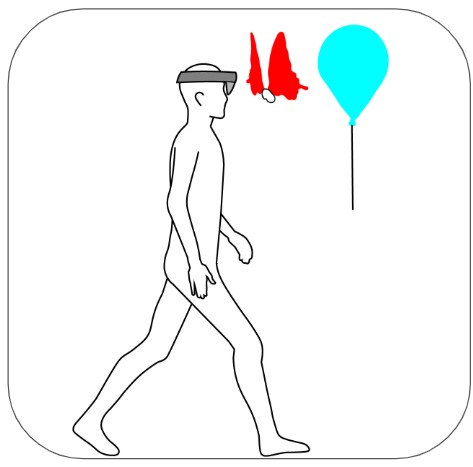}     \\ \bottomrule
\end{tabular}
\end{table}

\par
\subsubsection{Virtual Reality}
From the VR tasks, we extracted idle, pointing, button and physical interactions.
Specifically, we extracted time intervals in which participants were only looking at the robot while it was executing either a pick or a place automation. Those time frames were considered as idle actions, as participants were only looking at the scene without interacting with any of the virtual contents. Idle intervals in which participants were mentally summing the numbers for the arithmetic task were considered as high workload idles, while those cases in which participants were not engaged in the arithmetic task, nor in any interactions with virtual objects, were considered as low workload idles.
The pointing action was identified by selecting time periods in which participants were using the VR controller for pointing the numbers on the virtual keyboard shown in figure X. In that experimental phase, they where reporting the sum at the previously performed arithmetic task.

In the button interaction, participants guided the virtual robot through the pick-and-place task by only pressing specific buttons on the VR controller. Differently, in the physical interactions, participants physically touched the virtual robot and moved their own arm for relocating it over the worktable. In line with what demonstrated by the authors~\cite{nenna2022virtualization}, both button and physical interactions were categorized according to the level of workload involved. Specifically, when the pick-and-place was performed concurrently with the arithmetic task, participants were imposed with higher workload compared to when performing the pick-and-place task without additional tasks. Table~\ref{tab:VR_actions} represents the actions isolated in the VR environment.

\begin{table}[ht!]
\caption{Virtual Reality actions organized per type of operation and workload level.}
\label{tab:VR_actions}
\centering
\begin{tabular}{c| >{\centering\arraybackslash}m{2cm}  >{\centering\arraybackslash}  >{\centering\arraybackslash}m{2cm}  >{\centering\arraybackslash}m{2cm}  >{\centering\arraybackslash}m{2cm}} \toprule     
\multirow{2}{*}{\diagbox{\textit{Workload}}{ \textit{Operation}} }  & \multirow{2}{*}{\textbf{\textit{Idle}}} & \multirow{2}{*}{\textbf{\textit{Pointing}}} & \textbf{\textit{Button }}  & \textbf{\textit{Physical}} \\ 
    & & & \textbf{\textit{Interaction}} & \textbf{\textit{Interaction}}\\
   
    \toprule
  \textit{\textbf{Low}}
& \includegraphics[width = 2cm]{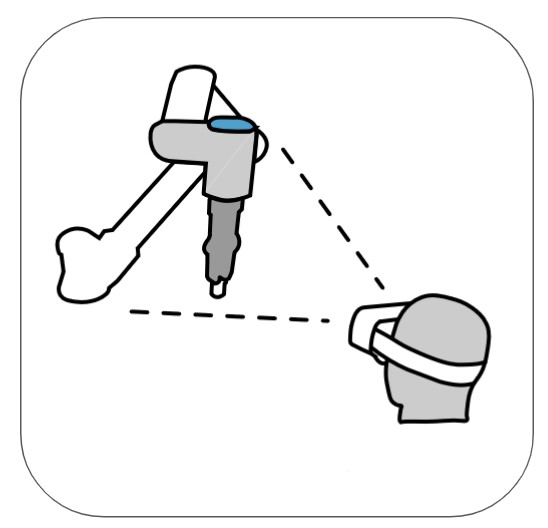}  & \includegraphics[width=2cm]{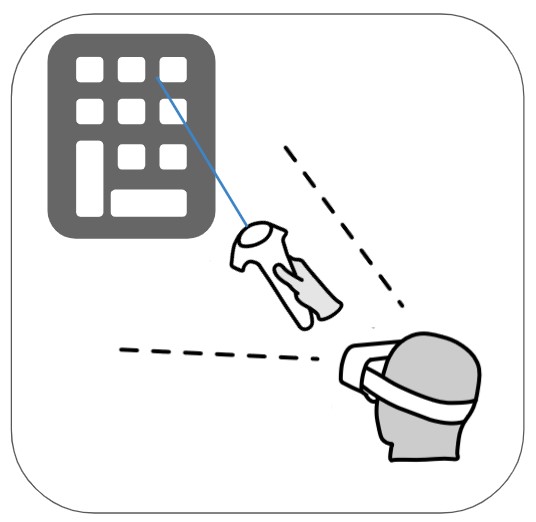} & \includegraphics[width=2cm]{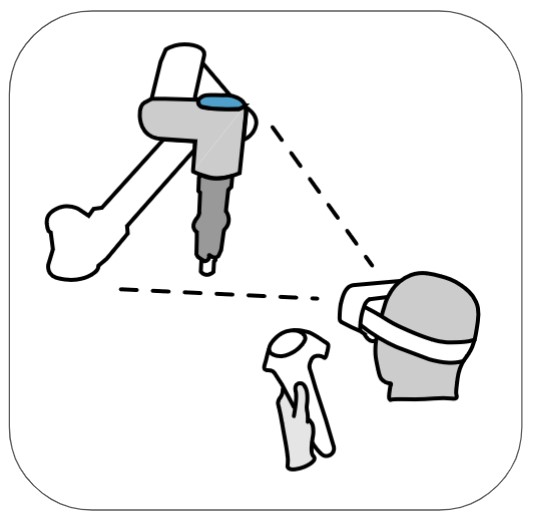} &    
  \includegraphics[width=2cm]{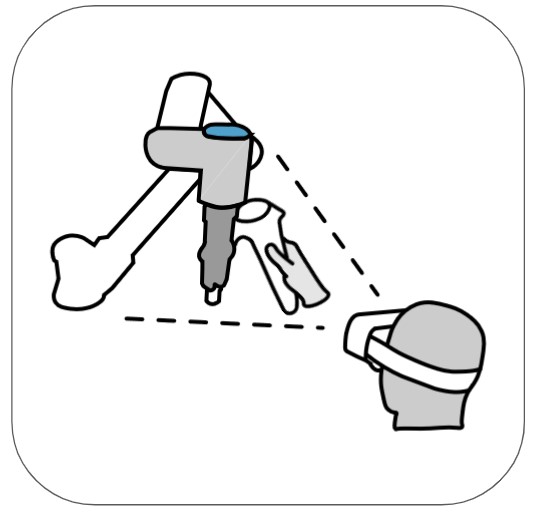} \\
  \textit{\textbf{High}} &  \includegraphics[width=2cm]{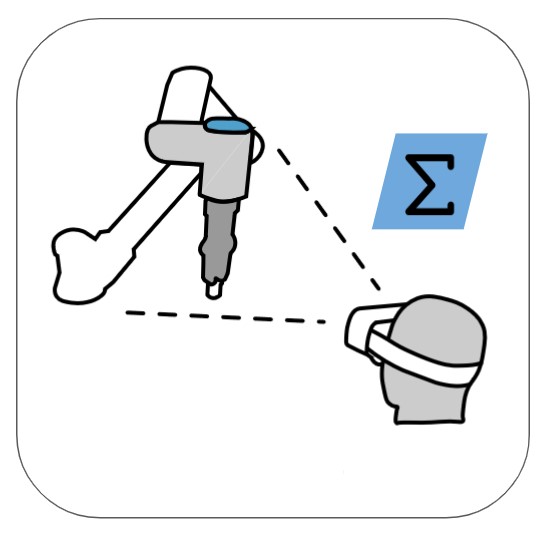}   & -- &    \includegraphics[width=2cm]{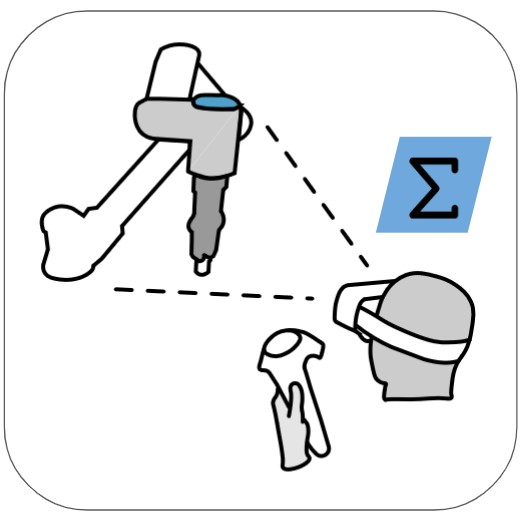} & \includegraphics[width=2cm]{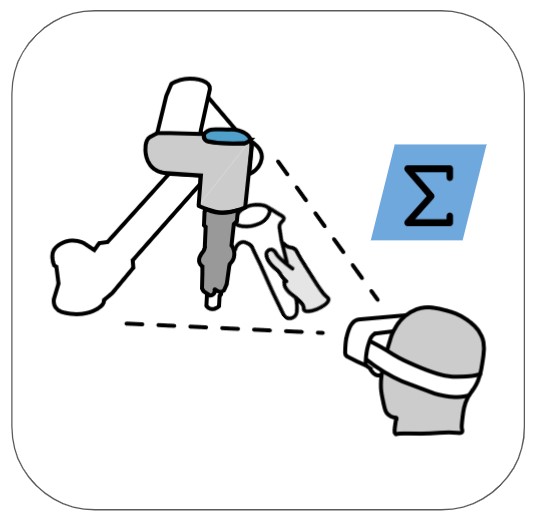}    \\ \bottomrule
\end{tabular}
\end{table}

\subsection{Implementation}\label{ssec.experiment/implementation}

\subsubsection{De-biasing and Feature Extraction}
AR and VR datasets contain different type of raw features acquired from the sensors.
We now describe, for each category of sensors, a description of the features and de-biasing techniques we applied.
\begin{itemize}
    \item \feature{Head Position} (AR and VR), represented as a 3D coordinate (x, y, z) measuring the relative distance (in meters) of the user from a center point in the virtual environment. This feature might contain both sessions and users static traits (e.g., height). We thus derived different variants of this information, such as the movement, computed as the norm between two points at 5 timestamp of distance, and the vertical oscillation, computed as the difference between two height values at 5 timestamp of distance.
    \item \feature{Head Rotation} (AR and VR), represented as a 3D value. For each axis, we compute its angular speed by considering points at 5 timestamp of distance. This transformation can remove information related to trials (e.g., specific positioning of objects with respect to the participant).
    \item \feature{Eyes} (VR), includes data on pupil size (in millimeters) and eye openness (0-1), for both left and right eyes. 
    It is to notice that, in order to overcome possible confounding variables~\cite{kramer2020physiological} \cite{mathot2018pupillometry}, it is usually appropriated to preprocess the raw eye data for flattening individual differences. 
    However, as the aim of the present work was specifically to capture individual traits and behaviors for allowing identification/profiling, we opted for not preprocessing eye-tracking data. 
    On the contrary, we leveraged the individual differences in pupil size and eye openness~\cite{bargary2017individual} \cite{fawcett2022individual} \cite{aminihajibashi2019individual} for better identifying and profiling users. 
    Further, we enhance this set of features by additionally computing the symmetry among the eyes for both pupil dilatation and eye openness.
    On an applied level, using the raw output of the HTC Vive Pro Eye device speeds up the identification/profiling process and allows higher generalizability to multiple VR devices.
    \item \feature{Controller Position} (VR), represented as a 3D coordinates (x, y, z) relative to the virtual environment center point. Similarly to the \feature{head position}, this feature might contain both sessions and users traits. We thus transform it in the movement, computed as the norm between two points at 5 timestamps of distance.
    \item \feature{Controller Rotation} (VR) represented as a 3D value. We conduct the same process of \feature{head rotation}.
\end{itemize}

Finally, each feature of the previously describe families is aggregated with tsfresh\footnote{\url{https://tsfresh.readthedocs.io/en/latest/index.html}}. Given a time series, this library extracts more than 100 features, including average, standard deviation, quantile, and entropy. We further refined the features by keeping only the relevant ones.\footnote{We used tfresh feature\_selection function: \url{https://tsfresh.readthedocs.io/en/latest/api/tsfresh.feature_selection.html}} Thus, starting from the raw time series of a single action within a single tasked performed in a single trial by a single user, we extract a single aggregated data point. The process is repeated for all the users, trials, actions, and tasks, obtaining 9360 datapoints in AR, and 16520 datapoints in VR. 

\subsubsection{Models Training and Validation}
In our experiments, we test four different algorithms: logistic regression, ridge classifier, decision tree, and random forest. 
As a baseline, we defined a Dummy classifier that randomly predicts the outcome based on the training ground-truth distribution.
For each experiment presented in Section~\ref{sec.results}, we adopt a common validation strategy: for each discussed model, we find the best hyper-parameters through a grid-search validation based on a training, validation and testing split of 70\%, 10\%, and 20\% of samples, respectively. 
For private inferring tasks (i.e., age and gender), the splits contain different set of users, i.e., users in training are not present in validation and testing set. Similar, users in validation are not present in both training and testing set. 
Machine learning models are designed as a multilabel task for the user identification task. On the opposite, we considered binary tasks both age (i.e., young and old) and gender (i.e., male and female). Note that the young class correspond to users defined in $[19 - 24]$ (AR) and $[23 - 30]$; the old class is defined in $[25, 29]$ (AR) and $[31 - 69]$.
We now report the parameter grids involved in the grid-searches. 
\begin{itemize}
    \item Logistic Regression (LR). C: ${0.1, 1, 10}$.
    \item Ridge (RI). Alpha: ${0.01, 0.1, 1., 10}$. Fit intercept: ${False, True}$.
    \item Decision Tree (DT). Max Depth: ${3, 5, 7}$. Min samples leaf: $1, 3, 5$.
    \item Random Forest (RF). N estimators: ${50, 100, 150}$. Max Depth: ${3, 5, 7}$. Min samples leaf: $1, 3, 5$.
\end{itemize}
To provide accurate results, each experiment is repeated five times. We thus report both mean and standard deviation of the F1-scores (with macro average).
We implemented our experiments in Python 3.8.5 and we used Scikit-Learn~\cite{scikit-learn} library for training models and validation algorithms. 

\section{Results}\label{sec.results}
In this section, we present the results of our experiments. 
We present both results for the task and action levels, in sections~\ref{ssec.experiment/task} and ~\ref{ssec.experiment/action}, respectively.
We then conclude with an ablation study to better understand the effect of different sensors to models' performance (Section~\ref{ssec.results/ablation}),

\subsection{Task-Level}\label{ssec.results/task}
In this section, we present profiling performance at a task-level. In particular, each presented experiment consider distinctly the tasks presented in Section~\ref{ssec.experiment/task}. In more details, we train, validate, and test our model only on the task under investigation, predicting each time the identity, age, and gender separately. For instance, we train a specific model to predict gender based only on the Mental Task.
\subsubsection{Identification} 
\begin{figure}[h!]
\centering
\begin{subfigure}{\textwidth}
    \includegraphics[width=\textwidth]{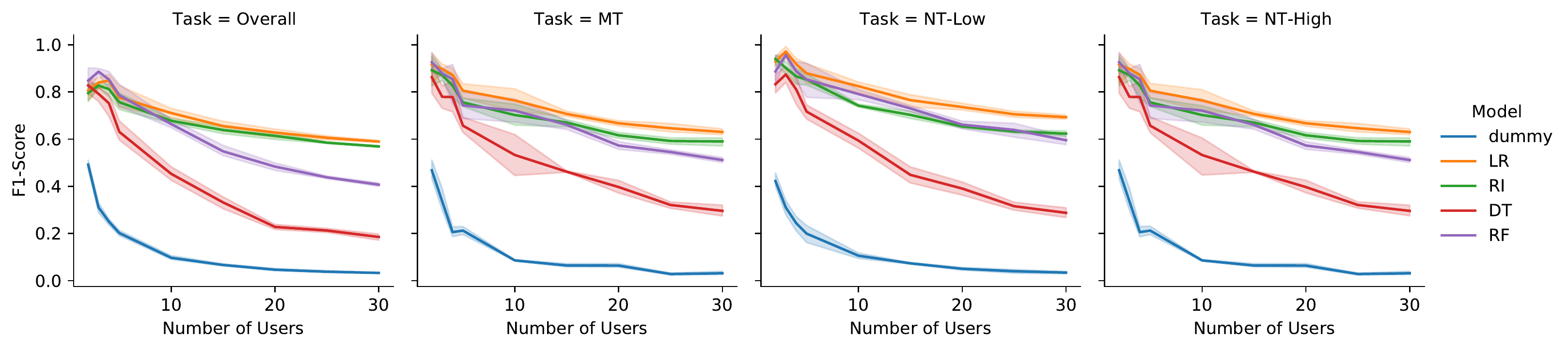}
    \caption{Augmented Reality}
    \label{fig:AR_task_id}
\end{subfigure}
\par\bigskip
\begin{subfigure}{\textwidth}
    \includegraphics[width=\textwidth]{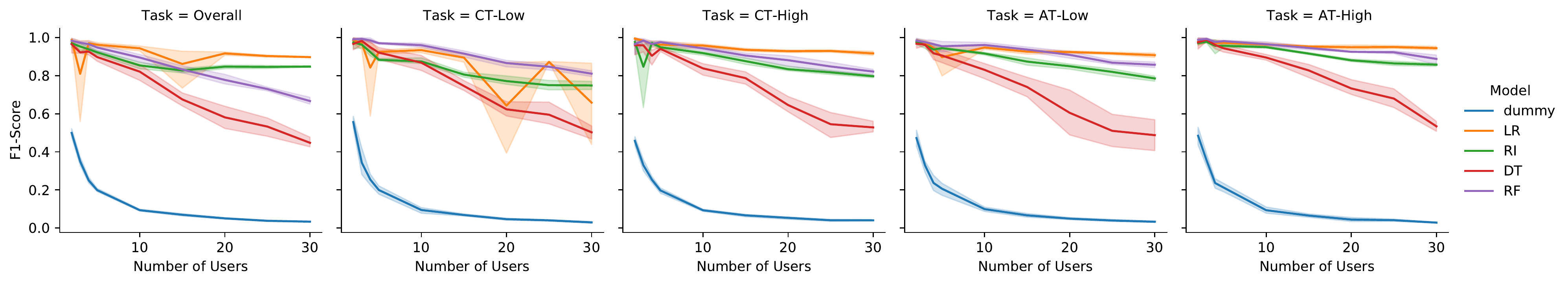}
    \caption{Virtual Reality}
    \label{fig:VR_task_id}
\end{subfigure}

\caption{User Identification on task-level.}
\label{fig:AR_VR_Id}
\end{figure}

Figure~\ref{fig:AR_VR_Id} shows the identification results in AR and VR environments.
LR and RI achieved the highest (and comparable) performances in AR, whereas LR and RF performed best in VR. 
In general, all our algorithms outperform the baseline (Dummy).
Looking at the results on the Overall Tasks, both in VR (OT-VR) and AR (OT-AR), we immediately notice that in VR identification, the performances remain quite stable as the number of users increases, while AR degrades significantly. Indeed, AR best algorithms performance goes from near 90\% F1-Score (two users) to slightly above 60\% F1-Score (30 users). Instead, in VR, LR yields almost perfect prediction on two users, while the F1-Score is above 95\% when performing identification over 30 users. This might reflect the different amount of sensors available in VR (headset, controller, and eye-related behaviors) compared to those available in AR (only headset-related behaviors). We further discuss the impact of each of the involved sensors in Section~\ref{ssec.results/ablation}.
\par
When looking at the individual tasks, we can see that the identification algorithm performs even better than in the overall task, particularly in AR. For instance, we reached 70\% F1-Score over 30 users in the NT-Low, which is roughly 10\% higher than in the OT-AR. One reason of this result might be related to the nature of the performed task: in the NT-Low, participants were actively moving in the surroundings without performing any additional task. Therefore, their movements might have been more linear compared to the situation in which they performed the same task under high workload (NT-High), thus revealing more identifiable movements' patterns.
The same does not apply to the VR scenario. Here, when looking at each of the identified actions, the higher the workload the better the performance of the identification algorithm. Indeed, the best performance was obtained at the AT-High and CT-High, where the F1-Score was around 95\% and 97\%, respectively. Again, possible explanations might be related to the nature of the tasks and to the number of sensors embedded in the devices. In the VR scenario, participants were only moving their upper body, and in the high workload conditions they were additionally engaged in a secondary mental task. We know from literature that higher workload is related to higher changes in the eye behavior \cite{nenna2022virtualization}. Therefore, the VR-embedded eye-tracker might have had an important impact on the identification performance, particularly when users were under higher mental strain rather than when performing less demanding tasks (i.e., CT-Low, AT-Low).



\subsubsection{Age}
\begin{figure}[h!]
\centering
\begin{subfigure}{0.4\textwidth}
    \includegraphics[width=\textwidth]{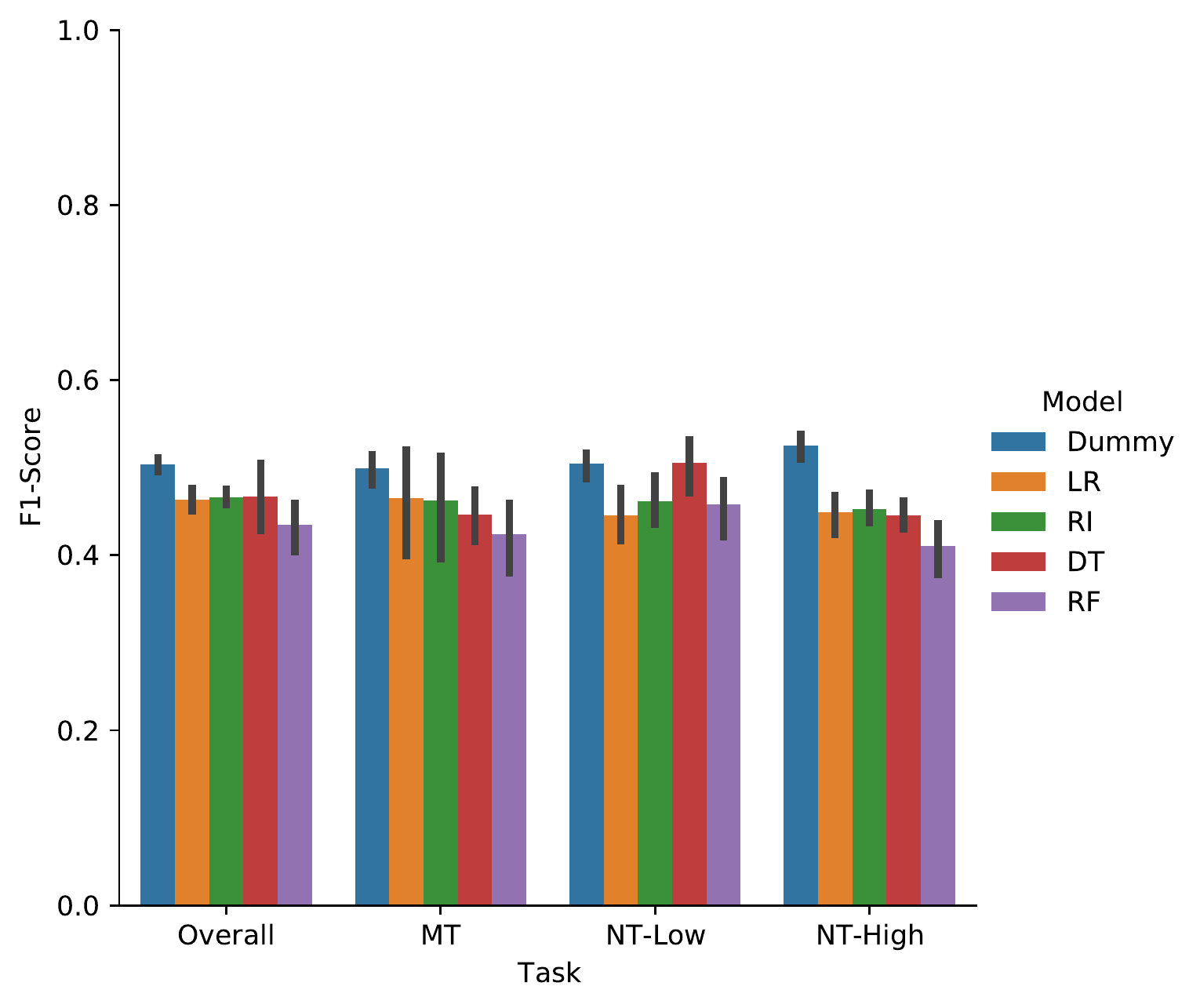}
    \caption{Augmented Reality.}
\end{subfigure}
\hspace{1cm}
\begin{subfigure}{0.4\textwidth}
    \includegraphics[width=\textwidth]{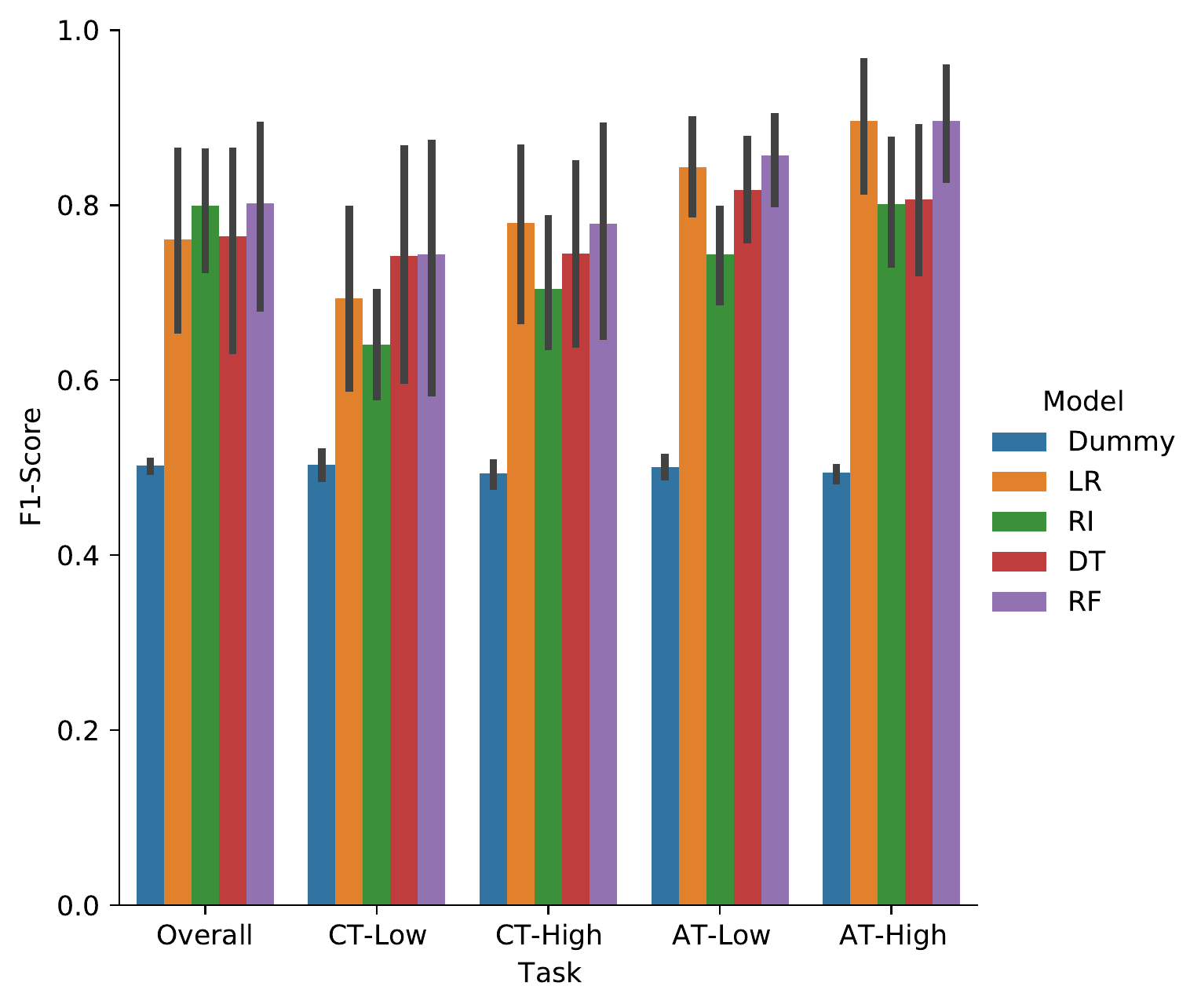}
    \caption{Virtual Reality.}
\end{subfigure}
\caption{Age profiling on task-level.}
\label{fig:task-age}
\end{figure}

Figure~\ref{fig:task-age} shows the age classification results in AR and VR environment at task-level. 
Results from the age profiling clearly yielded better performance in the VR compared to the AR scenario. While in VR all models performed significantly better than the baseline, in AR the F1-Score was consistently lower than the baseline, in all tasks. This is likely to be related to the low age variability of participants that took part in the AR experiment. Therefore, we here only discuss performances of age profiling only in relation to the VR experiment.
\par

In VR, the LR and RF algorithms appear to perform better then the other models in all tasks, but in the OT, whereas RI produced higher F1-Score compared to LR. On the task-level, the users' age was profiled with higher accuracy when they performed the pick-and-place task via physical actions (AT-High and AT-Low, in which F1-Score was around 90\% and 85\% respectively) compared to controller buttons (CB-High, CB-Low, in which F1-Score was below 80\% in both cases). A possible interpretation on this point is that the movements' pattern of older users might have been quite different from younger users. Also, we know from literature that robot teleoperation is significantly influenced by age~\cite{grabowski2021teleoperated}. In this view, our algorithm was particularly successful in detecting users' age during the pick-and-place task only when physical actions were allowed.

\subsubsection{Gender}
\begin{figure}[h!]
\centering
\begin{subfigure}{0.4\textwidth}
    \includegraphics[width=\textwidth]{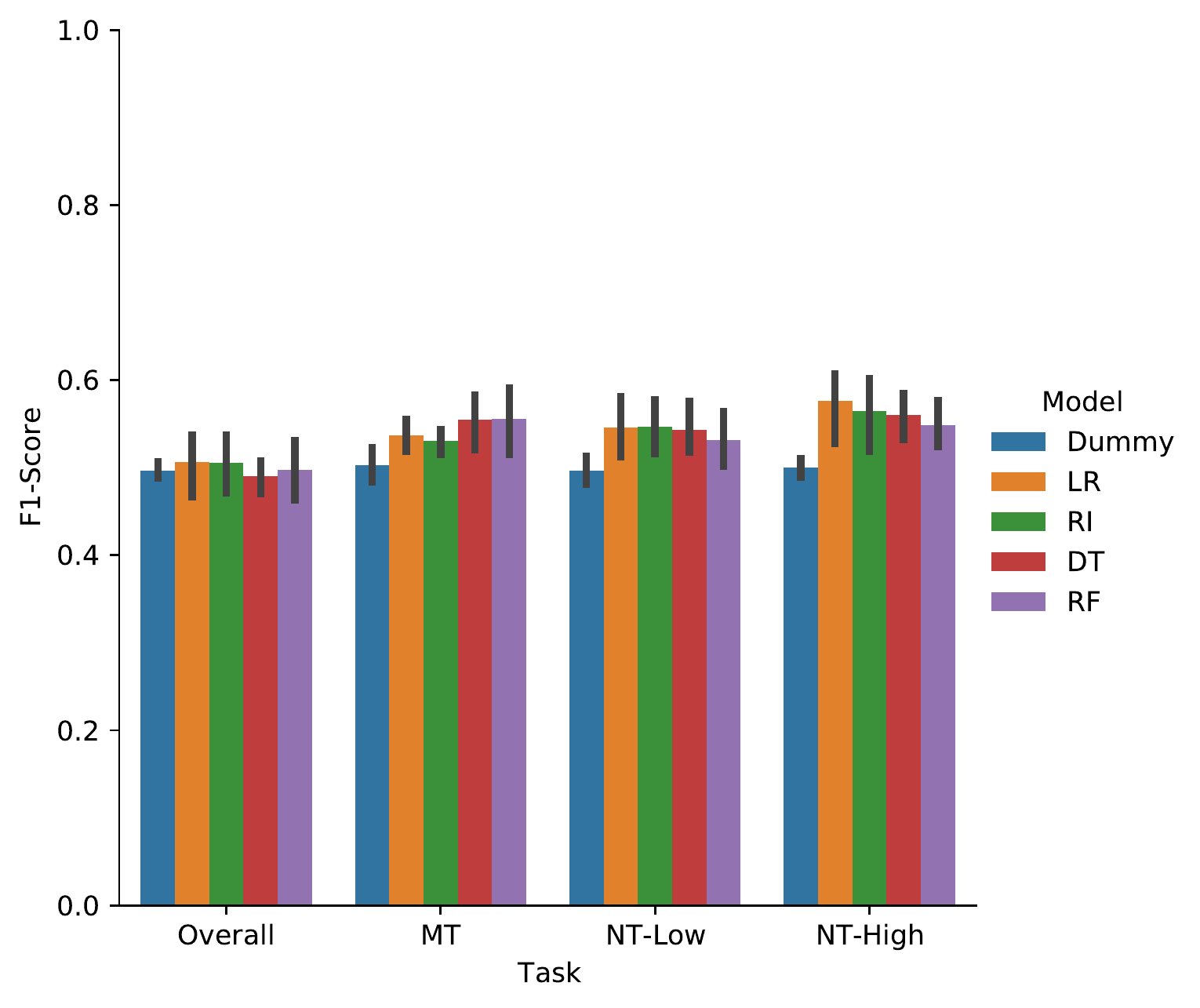}
    \caption{Augmented Reality.}
\end{subfigure}
\hspace{1cm}
\begin{subfigure}{0.4\textwidth}
    \includegraphics[width=\textwidth]{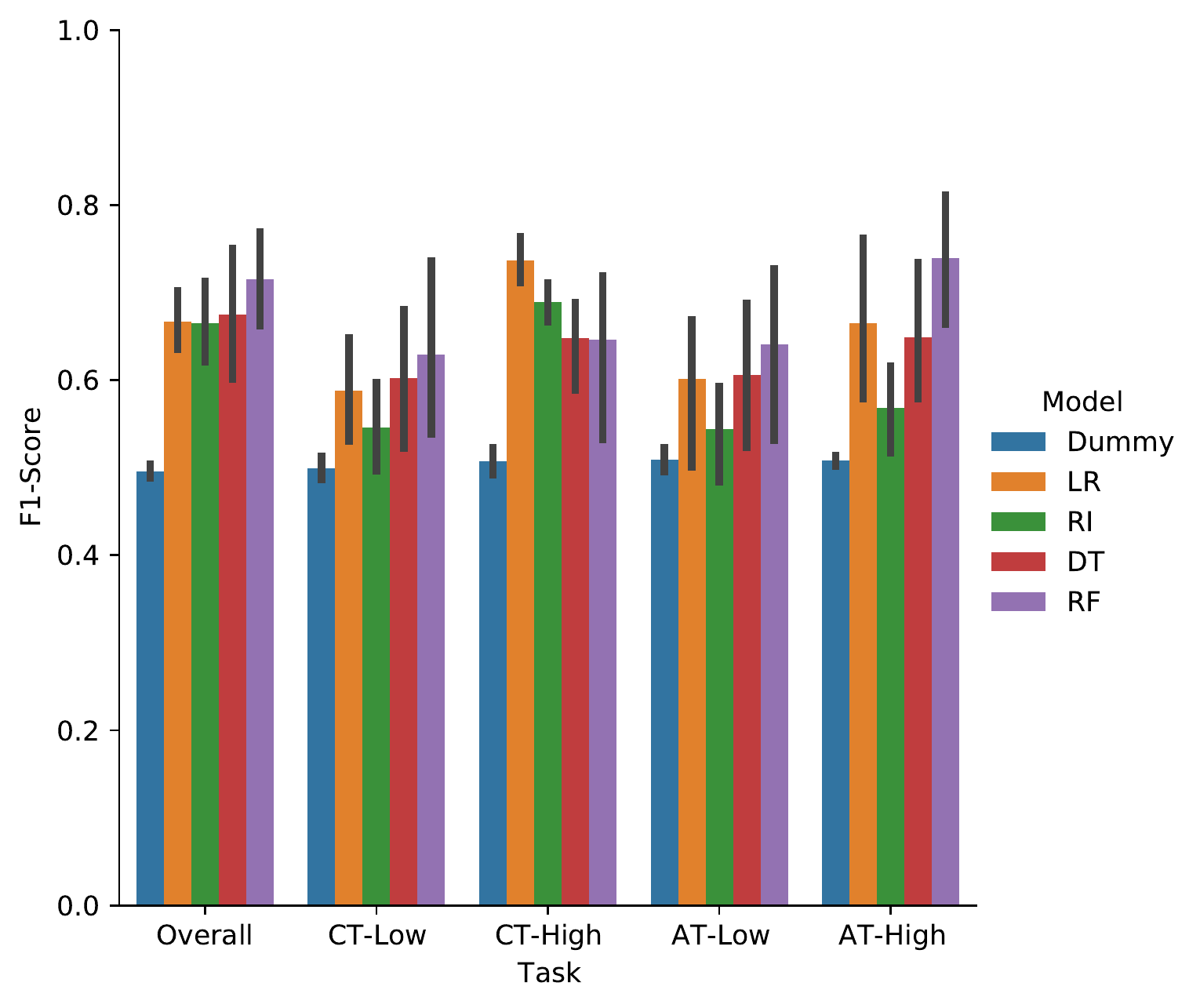}
    \caption{Virtual Reality.}
\end{subfigure}
\caption{Gender profiling on task-level.}
\label{fig:task-gender}
\end{figure}

Figure~\ref{fig:task-gender} shows the gender classification results in AR and VR environment at task-level. 
When profiling users' gender, we obtained substantially better results in VR compared to AR. 
Indeed, in VR, all the tested algorithms performed above the baseline (dummy). More specifically, we can observe a better performance obtained through LR and RF, which reached a maximum F1-Score of 75\%. Differently, when detecting users' gender in the AR scenario, our algorithms performed only 5-10\% above the baseline.
\par
For the algorithms' performance within each of the identified tasks in VR, a better performance is achieved in tasks involving higher workload (CT-High, AT-High) compared to those under low workload (CT-Low, AT-Low). These results align with recent literature on behavioral gender differences in the VR pick-and-place task. For instance, Nenna et al.~\cite{nenna2022influence} demonstrated how men outperformed women in the pick-and-place tasks in terms of task execution time, particularly when using controller buttons. These differences might have been even more marked when performing an additional mental task, thus allowing a better gender profiling.
We observe a similar trend in the AR scenario, in which better performance are reached in the task involving higher workload (NT-High). This behavior reflects previous findings related to the different walking pattern between men and women~\cite{nenna2021augmented}. Indeed, on average, the walking velocity of men is significantly higher than women' one, particularly under high workload. As we were recording the headset shifts in time, the different walking velocity might have been prominent in the gender profiling.

\subsection{Action-Level}\label{ssec.results/action}
Starting from the results obtained in the overall task, we here aimed at seeing whether some actions had a particular influence on the identification and profiling performances. Specifically, we opted for leveraging the model that demonstrated better results, which was the Logistic Regression (LR). Each presented experiment consider distinctly the tasks presented in Section~\ref{ssec.experiment/action}. In more details, we train, validate, and test our model only on the action under investigation, predicting each time the identity, age, and gender separately. For instance, we train a specific model to predict age based only on Button Interaction with Low Workload.

\subsubsection{Identification}
Table~\ref{tab.action-identification} shows the identification results in AR and VR environment at action-level.  
Performance obtained on task-level reached an F1-Score of about 60\% in the AR and above 90\% in the VR scenario. 
When looking at the action-level, specifically for AR, we see that the walking action reaches the highest performance (F1-Score is about 0.80\% under low workload and 0.78\% under high workload), while the search action and button interaction reveal F1-Scores below 0.70\%. This suggests that the walking action is prominent in identifying users in AR, possibly because the walking pattern is the most singular feature in such a use-case of AR.
Differently, in VR, we observe higher F1-Scores for both button and physical interactions specifically under high workload (F1-Score is about 0.96\% in both cases). Also the pointing action reached a very similar F1-Score (0.96\%), while the idle time intervals yield lower F1-Scores (below 0.80\% both under high and low workloads). It seems that the most interactive actions (using controller buttons, pointing and physically moving the upper body) thus yield better results compared to periods in which users were passively looking at the virtual surroundings.


\begin{table}[!ht]
\caption{User identification on action-level organized per type of operation and workload level. Random guess at 0.03 for both AR and VR tasks. All the measures in F1-Score.}
\centering
\begin{tabular}{c|ccc|cccc} \cmidrule{2-8}
    & \multicolumn{3}{c|}{\textit{Augmented Reality}} & \multicolumn{4}{c}{\textit{Virtual Reality}}\\ \cmidrule{1-8}
     \multirow{2}{*}{\diagbox{\textit{Workload}}{ \textit{Operation}} } & \textbf{\textit{Button}} & \multirow{2}{*}{\textbf{\textit{Search}}} & \multirow{2}{*}{\textbf{\textit{Walk}}}  & \multirow{2}{*}{\textbf{\textit{Idle}}} & \multirow{2}{*}{\textbf{\textit{Pointing}}} & \textbf{\textit{Button }}  & \textbf{\textit{Physical}} \\ 
    & \textbf{\textit{Interaction}} & & & & & \textbf{\textit{Interaction}} & \textbf{\textit{Interaction}}\\ \toprule
\textit{\textbf{Low}}  &      --             &    \res{0.66}{0.03}     &  \res{0.80}{0.02}     &  \res{0.78}{0.02}    & \res{0.96}{0.01} & \res{0.92}{0.01} & \res{0.93}{0.02} \\
\textit{\textbf{High}} & \res{0.61}{0.02}  &  \res{0.69}{0.01}  &    \res{0.78}{0.02}   &  \res{0.86}{0.01}    & -- & \res{0.96}{0.00} & \res{0.96}{0.01} \\ \bottomrule
\end{tabular}
\label{tab.action-identification}
\end{table}

\subsubsection{Age}
Table~\ref{tab.action-age} shows the age classification results in AR and VR environment at action-level.  
Users' age was profiled with an F1-Score of about 0.5\% on the overall task executed in AR, and 0.80\% in VR. As the age profiling revealed to be unsuccessful in AR, we will not pay close attention on the action-level results on this use-case. 
This results confirm what we observed at task-level (see Figure~\ref{fig:task-age}).
Regarding the VR scenario, we can note that, under low workload, the pointing (F1-Score = 0.88\%) and physical interactions (F1-Score = 0.82\%) were the most crucial in profiling users' age, compared to actions allowing less interactivity with the virtual environment (F1-Scores below 0.80\%). This might be an hint for a different movement and interaction pattern shown by older and younger users, which comes out particularly when higher freedom of movement is allowed. This is also in line with what observed on task-level. Moreover, this trend becomes even more evident when the physical interactions are performed under high workload (F1-Score = 0.90\%), likely reflecting the multitasking and motor difficulties related to age~\cite{li2005ecological}.

\begin{table}[!ht]
\caption{Age profiling on action-level organized per type of operation and workload level. Random guess at 0.5 for both AR and VR tasks. All the measures in F1-Score.}
\centering
\begin{tabular}{c|ccc|cccc} \cmidrule{2-8}
    & \multicolumn{3}{c|}{\textit{Augmented Reality}} & \multicolumn{4}{c}{\textit{Virtual Reality}}\\ \cmidrule{1-8}
     \multirow{2}{*}{\diagbox{\textit{Workload}}{ \textit{Operation}} } & \textbf{\textit{Button}} & \multirow{2}{*}{\textbf{\textit{Search}}} & \multirow{2}{*}{\textbf{\textit{Walk}}}  & \multirow{2}{*}{\textbf{\textit{Idle}}} & \multirow{2}{*}{\textbf{\textit{Pointing}}} & \textbf{\textit{Button }}  & \textbf{\textit{Physical}} \\ 
    & \textbf{\textit{Interaction}} & & & & & \textbf{\textit{Interaction}} & \textbf{\textit{Interaction}}\\
    \toprule
\textit{\textbf{Low}}  &  --  & \res{0.40}{0.03} & \res{0.45}{0.02}    & \res{0.77}{0.10} & \res{0.88}{0.06} & \res{0.70}{0.09} & \res{0.82}{0.05} \\
\textit{\textbf{High}} & \res{0.47}{0.02}  &  \res{0.44}{0.01}  &    \res{0.49}{0.02}   &  \res{0.83}{0.09}    & -- & \res{0.81}{0.07} & \res{0.90}{0.05} \\ \bottomrule
\end{tabular}
\label{tab.action-age}
\end{table}

\subsubsection{Gender}
Table~\ref{tab.action-gender} shows the gender classification results in AR and VR environment at action-level.  
On task-level, our algorithms generated an F1-Score of about 0.5\% in AR and above 0.7\% in VR. Even though the gender profiling did not perform sufficiently well in AR, we can here observe that, under high workload, both walk (F1-Score = 0.6\%) and search (F1-Score = 0.58\%) had a major influence in detecting the user gender compared to the same actions performed under low workload, and also to the button interaction (all f-scores < 0.50\%). These results are in line with what observed on task-level, whereby the gender profiling performed better in the NT-high compared to NT-low. Additionally, we here observe how the walking action has the larger influence on the accuracy of gender profiling compared to the other actions. Again, it might be related to different walking velocities demonstrated by men and women, particularly under high workload \cite{nenna2021augmented}.
\par
When looking at the actions performed in VR, the pointing action stands out. With an f-score of 0.82\%, it strongly contributes to the gender profiling compared to all other actions. This might be related both to a singular movement pattern and/or to gender-related eye parameters' variations.
Further, results obtained at task-level on a better performance achieved under high compared to low workload are here confirmed only for button interactions. Indeed, the F1-Score at button interactions is about 0.08\% higher when users are under high rather than low workload. Again, this reflects results showed in previous study demonstrating faster operation times in men compared to women specifically when using controller buttons, but not when acting via physical actions~\cite{nenna2022virtualization}. This suggests that profiling users' gender might be easier during tasks involving button interactions, but not in those allowing higher interactivity with the virtual environment.

\begin{table}[!ht]
\caption{Gender profiling on action-level organized per type of operation and workload level. Random guess at 0.5 for both AR and VR tasks. All the measures in F1-Score.}
\centering
\begin{tabular}{c|ccc|cccc} \cmidrule{2-8}
    & \multicolumn{3}{c|}{\textit{Augmented Reality}} & \multicolumn{4}{c}{\textit{Virtual Reality}}\\ \cmidrule{1-8}
     \multirow{2}{*}{\diagbox{\textit{Workload}}{ \textit{Operation}} } & \textbf{\textit{Button}} & \multirow{2}{*}{\textbf{\textit{Search}}} & \multirow{2}{*}{\textbf{\textit{Walk}}}  & \multirow{2}{*}{\textbf{\textit{Idle}}} & \multirow{2}{*}{\textbf{\textit{Pointing}}} & \textbf{\textit{Button }}  & \textbf{\textit{Physical}} \\ 
    & \textbf{\textit{Interaction}} & & & & & \textbf{\textit{Interaction}} & \textbf{\textit{Interaction}}\\
    \toprule
\textit{\textbf{Low}}  &  --  & \res{0.50}{0.02} & \res{0.45}{0.06}    & \res{0.60}{0.10} & \res{0.82}{0.09} & \res{0.62}{0.05} & \res{0.66}{0.11} \\
\textit{\textbf{High}} & \res{0.54}{0.03}  &  \res{0.58}{0.03}  &    \res{0.60}{0.06}   &  \res{0.63}{0.05}    & -- & \res{0.74}{0.06} & \res{0.66}{0.08} \\ \bottomrule
\end{tabular}
\label{tab.action-gender}
\end{table}

\subsection{Sensors Relevance - Ablation Study}\label{ssec.results/ablation}
In this section, we conduct an ablation study to understand which sensors contribute the most in our identification, age, and gender predictions. In brief, we trained a Logistic Regression (LR) using only specific subsets of features. In the AR environment, we distinguish between \feature{Head Position} and \feature{Head Rotation} features. In VR, we also consider \feature{Eyes}, \feature{Controller Position}, and \feature{Controller Rotation} features. The ablation study was carried out both at Task-Level (Section~\ref{subsub:abl-task}) and Action-Level (Section~\ref{subsub:abl-act}). 

\subsubsection{Task-level}
\label{subsub:abl-task}

Table~\ref{tab:AR_task_abl} and Table~\ref{tab:VR_task_abl} show the results of the ablation study for AR and VR tasks, respectively. 
In the AR environment, \feature{Head Rotation} features are predominant in the Mental Task for identification and gender prediction. Indeed, in this task, participants were standing still and were instructed to don't move their head; however, it was plausible that their head oscillated in singular ways, which were detected by our algorithm and leveraged for their identification. In opposition, during the navigation task, \feature{Head Position} has more impact in all the targets, given that it records the walking patterns. Such pattern was used in the literature to identify people~\cite{katiyar2013study}, and could help in Age and Gender prediction as well.  

In VR, the identification stage seems to be driven manly by \feature{Eyes} features, followed by \feature{Controller} features. Reasonably, eyes blinking patterns and pupils' dilatation can be person-specific~\cite{bargary2017individual} \cite{fawcett2022individual} \cite{aminihajibashi2019individual}, and thus act as a biometric feature. The controllers, instead, were the main means to interact with the virtual world. Thus, it is reasonable that how a person interact within the environment helps in the identification. This result is in line with recent founds on video games using mouse and keyboards to profile users~\cite{conti2020pvp}. Therefore, we could expect AR identification achieve better performances if such sensors are available, particularly eyes trackers, as reasoned before in Section~\ref{ssec.results/task}. In predicting the age, the \feature{Controller} features yields the best performance. This finding can be the consequence of younger people being more familiar with joystick usage. When the workload is high, younger participants may pay more attention to the task rather than how to use the joystick. Moreover, in a low workload scenario, \feature{Head} and \feature{Eyes} features contributes similarly. 
On the other hand, in gender inference, the \feature{Head} and \feature{Eyes} features play the most significant role. Indeed, as shown in past literature, there are gender-based differences in how they visually explore a virtual world~\cite{sargezeh2019gender}. \feature{Controller} features influence the prediction mainly in high workload controller based tasks.

\begin{table}[h!]
\scriptsize
\caption{Ablation study of sensor importance at task-level in AR. All the measures in F1-Score.}
\centering
\label{tab:AR_task_abl}
\begin{tabular}{l|l|ccc}
\cmidrule{2-5}
\multicolumn{1}{l}{} &  & \textit{\textbf{Identification}} & \textit{\textbf{Age}} & \textit{\textbf{Gender}} \\\cmidrule{2-5}
\multicolumn{1}{l}{} & \textbf{Guessing} & 0.03 & 0.5 & 0.5 \\\cmidrule{2-5}
\multicolumn{1}{l}{} & \textbf{Mental Task} &  &  &  \\
\multicolumn{1}{l}{} & Head Position & 0.38 & \textbf{0.46} & 0.51 \\
\multicolumn{1}{l}{} & Head Rotation & \textbf{0.54} & 0.40 & \textbf{0.55} \\\midrule
\multirow{3}{*}{\rotatebox[origin=c]{90}{Low W.}} & \textbf{Navigation Task} & \textbf{} &  &  \\
 & Head Position & \textbf{0.64} & \textbf{0.45} & \textbf{0.56} \\
 & Head Rotation & 0.46 & 0.40 & 0.45 \\\midrule
\multirow{3}{*}{\rotatebox[origin=c]{90}{High W.}} & \textbf{Navigation Task} & \textbf{} &  &  \\
 & Head Position & \textbf{0.65} & \textbf{0.45} & 0.51 \\
 & Head Rotation & 0.48 & 0.44 & \textbf{0.52} \\\bottomrule
\end{tabular}
\end{table}

\begin{table}[h!]
\scriptsize
\caption{Ablation study of sensor importance at task-level in VR. All the measures in F1-Score.}
\centering
\label{tab:VR_task_abl}
\begin{tabular}{l|l|ccc}
\cmidrule{2-5}
 &  & \textbf{\textit{Identification}} & \textbf{\textit{Age}} & \textbf{\textit{Gender}} \\
\cmidrule{2-5}
 & \textbf{Guessing} & 0.03 & 0.5 & 0.5 \\
\cmidrule{2-5}
\multirow{12}{*}{\rotatebox[origin=c]{90}{Low Workload}} & \textbf{Controller Based Task} &  &  &  \\
 & Head Position & 0.41 & 0.68 & \textbf{0.64} \\
 & Head Rotation & 0.45 & \textbf{0.76} & 0.55 \\
 & Eyes & \textbf{0.83} & 0.75 & 0.59 \\
 & Controller Position & 0.39 & 0.69 & 0.57 \\
 & Controller Rotation & 0.59 & 0.69 & 0.58 \\
 & \textbf{Action Based Task} &  &  &  \\
 & Head Position & 0.50 & 0.76 & \textbf{0.62} \\
 & Head Rotation & 0.51 & 0.76 & 0.60 \\
 & Eyes & \textbf{0.83} & 0.74 & 0.54 \\
 & Controller Position & 0.51 & 0.76 & 0.58 \\
 & Controller Rotation & 0.68 & \textbf{0.81} & 0.55 \\
 \midrule
\multirow{12}{*}{\rotatebox[origin=c]{90}{High Workload}} & \textbf{Controller Based Task} &  &  &  \\
 & Head Position & 0.48 & 0.73 & 0.61 \\
 & Head Rotation & 0.56 & 0.68 & 0.57 \\
 & Eyes & \textbf{0.88} & \textbf{0.79} & \textbf{0.69} \\
 & Controller Position & 0.45 & 0.78 & 0.60 \\
 & Controller Rotation & 0.64 & 0.68 & 0.62 \\
 & \textbf{Action Based Task} &  &  &  \\
 & Head Position & 0.55 & 0.75 & 0.53 \\
 & Head Rotation & 0.55 & 0.80 & \textbf{0.62} \\
 & Eyes & \textbf{0.89} & 0.83 & \textbf{0.62} \\
 & Controller Position & 0.57 & 0.86 & 0.50 \\
 & Controller Rotation & 0.73 & \textbf{0.87} & 0.50 \\\bottomrule
\end{tabular}
\end{table}

\subsubsection{Action-level} \label{subsub:abl-act}
Table~\ref{tab:AR_act_abl} and Table~\ref{tab:VR_act_abl} reports the results of the ablation study for AR and VR Actions, respectively. In AR, the \feature{Head Position} has more impact than \feature{Head Rotation} in predicting our target actions, especially for the walk action. This is reasonable given that such sensor is mainly recording the users' walking speed. \feature{Head Rotation} becomes relevant in the Button Interaction action, in which the participants could just rotate their head, and is quite useful to distinguish between genders. As in previous results, the age was difficult to predict. The only case in which we surpass the baseline is in the Walk action with high workload, but the improvement is too small to reason about it. 

Looking at VR, we notice that \feature{Head Position} remains relevant to predict the gender, particularly in scenarios with low workload. However, most of the times, the \feature{Eyes} features are the main discriminant to predict our targets. In identification, \feature{Eyes} reached the highest F1-Score in six out of seven actions, suggesting that these features might be the main reason behind the higher identification performances in VR rather than AR. Further,  \feature{Eyes} are predominant in low workload scenarios to predict the users' age.  \feature{Controller} features are quite useful to infer the user's age especially in high workload actions, while only small differences appear in their usage from people of different genders. Regarding the identification task, the \feature{Controller Rotation} appears more useful than \feature{Controller Position}. Last, it is interesting to see how in the idle actions, the \feature{Eyes} play a significant role, particularly in the high workload scenario, in which were able to identify a person with 81\% of F1-Score.


\begin{table}[h!]
\caption{Ablation study of sensor importance at action-level in AR. All the measures in F1-Score.}
\centering
\label{tab:AR_act_abl}
\scriptsize
\begin{tabular}{l|l|ccc}
\cmidrule{2-5}
 &  & \textit{\textbf{Identification}} & \textit{\textbf{Age}} & \textit{\textbf{Gender}} \\\cmidrule{2-5}
 & \textbf{Guessing} & 0.03 & 0.5 & 0.5 \\ \midrule
\multirow{6}{*}{\rotatebox[origin=c]{90}{Low Workload}} & \textbf{Search} &  &  &  \\
 & Head Position & 0.\textbf{60} & 0.40 & 0.52 \\
 & Head Rotation & 0.51 & 0.40 & 0.\textbf{58} \\
 & \textbf{Walk} &  &  &  \\
 & Head Position & 0.\textbf{77} & 0.44 & \textbf{0.60} \\
 & Head Rotation & 0.55 & \textbf{0.47} & 0.49 \\\midrule
\multirow{9}{*}{\rotatebox[origin=c]{90}{High Workload}} & \textbf{Button Interaction} &  &  &  \\
 & Head Position & 0.38 & \textbf{0.46} & 0.52 \\
 & Head Rotation & \textbf{0.56} & 0.40 & \textbf{0.56} \\
 & \textbf{Search} &  &  &  \\
 & Head Position & \textbf{0.62} & 0.40 & \textbf{0.60} \\
 & Head Rotation & 0.52 & \textbf{ 0.43} & 0.57 \\
 & \textbf{Walk} &  &  &  \\
 & Head Position & \textbf{0.75} & \textbf{0.51} & \textbf{0.53} \\
 & Head Rotation & 0.55 & 0.43 & 0.47 \\\bottomrule
\end{tabular}
\end{table}


\begin{table}[h!]
\scriptsize
\caption{Ablation study of sensor importance at action-level in VR. All the measures in F1-Score.}
\centering
\label{tab:VR_act_abl}
\begin{tabular}{l|l|ccc}
\cmidrule{2-5}
 &  & \textit{\textbf{Identification}} & \textit{\textbf{Age}} & \textit{\textbf{Gender}} \\\cmidrule{2-5}
 & \textbf{Guessing} & 0.03 & 0.5 & 0.5 \\\midrule
\multirow{24}{*}{\rotatebox[origin=c]{90}{Low Workload}} & \textbf{Idle} &  &  &  \\
 & Head Position & 0.41 & 0.62 & \textbf{0.62} \\
 & Head Rotation & 0.44 & 0.69 & 0.59 \\
 & Eyes & \textbf{0.75} & \textbf{0.80} & 0.55 \\
 & Controller Position & 0.38 & 0.69 & 0.58 \\
 & Controller Rotation & 0.55 & 0.72 & 0.55 \\
 & \textbf{Pointer} &  &  &  \\
 & Head Position & 0.67 & 0.80 & 0.57 \\
 & Head Rotation & 0.73 & 0.83 & 0.62 \\
 & Eyes & \textbf{0.91} & \textbf{0.86} & \textbf{0.71} \\
 & Controller Position & 0.64 & 0.70 & 0.59 \\
 & Controller Rotation & 0.83 & 0.81 & 0.51 \\
 & \textbf{Button Interaction} &  &  &  \\
 & Head Position & 0.50 & 0.72 & \textbf{0.63} \\
 & Head Rotation & 0.55 & 0.73 & 0.56 \\
 & Eyes & \textbf{0.85} & \textbf{0.78} & 0.61 \\
 & Controller Position & 0.47 & 0.72 & 0.58 \\
 & Controller Rotation & 0.71 & 0.75 & 0.60 \\
 & \textbf{Physical Interaction} &  &  &  \\
 & Head Position & 0.59 & 0.75 & 0.62 \\
 & Head Rotation & 0.56 & 0.81 & \textbf{0.63} \\
 & Eyes & \textbf{0.87} & 0.80 & 0.61 \\
 & Controller Position & 0.63 & 0.74 & 0.57 \\
 & Controller Rotation & 0.75 & \textbf{0.85} & 0.56 \\\midrule
\multirow{18}{*}{\rotatebox[origin=c]{90}{High Workload}} & \textbf{Idle} &  &  &  \\
 & Head Position & 0.49 & 0.75 & 0.60 \\
 & Head Rotation & 0.47 & 0.76 & 0.55 \\
 & Eyes & \textbf{0.81} & 0.77 & \textbf{0.63} \\
 & Controller Position & 0.46 & \textbf{0.79} & 0.50 \\
 & Controller Rotation & 0.65 & \textbf{0.79} & 0.49 \\
 & \textbf{Button Interaction} &  &  &  \\
 & Head Position & 0.57 & 0.69 & 0.56 \\
 & Head Rotation & 0.65 & 0.66 & 0.55 \\
 & Eyes & \textbf{0.93} & \textbf{0.83} & \textbf{0.67} \\
 & Controller Position & 0.50 & 0.77 & 0.61 \\
 & Controller Rotation & 0.73 & 0.72 & 0.61 \\
 & \textbf{Physical Interaction} &  &  &  \\
 & Head Position & 0.63 & 0.82 & 0.54 \\
 & Head Rotation & 0.63 & 0.81 & 0.62 \\
 & Eyes & 0.71 & 0.87 & \textbf{0.66} \\
 & Controller Position & 0.66 & \textbf{0.90} & 0.45 \\
 & Controller Rotation & \textbf{0.79} & 0.86 & 0.49 \\\bottomrule
\end{tabular}
\end{table}
\section{Discussions and Conclusions} \label{sec.concl}
The profiling of users wearing virtual technologies can present several opportunities and threats, so it is important to examine it closely. 
In this work, we performed users' identification and profiling in two virtual-based scenarios, one involving AR and the other involving VR. 
We aimed to test different algorithms and leverage behavioral data outputted by the two virtual devices to accurately trace it back to the user identity and personal information (i.e., gender and age). 
Further, we developed a generic pipeline that can be used with different virtual devices and in different behavioral contexts: i.e., while walking, searching for landmarks in the surroundings, pointing to a virtual keyboard for typing, and operating on a virtual robot both via controller-based interaction and physical actions. 
Specifically, both virtual environments simulated highly realistic scenarios and most of these behaviors were executed both under high and low workload, giving a good insight on realistic applications of virtual technologies in the field.
\par

The results show that users can be identified and profiled both in AR and VR, with VR accuracy being higher. 
Specifically, in AR, user identification reached good results within the walking action at a low workload, while in VR, the identification algorithm was particularly successful when users performed more physical actions (i.e., pointing, physically interacting with the virtual robot) under a higher workload. 
As observed from the ablation study, this was mainly due to the additional eye-tracking sensors embedded in the VR but not in the AR headset. 
Indeed, while in VR the eye parameter had the most significant impact, the head movement revealed had the greatest influence on the AR users' identification. 

When detecting age, instead, our algorithms were not able to accurately detect the users' age in AR. 
This was plausibly related to the low age variability of the tested sample, as the age of participants that took part at the experiment ranged between 19 and 29. 
Differently, in VR, we worked on an experimental sample whose age ranged between 23 and 69 years old, and our algorithms were thus able to detect the user age with good accuracy. 
Age detection performed better in the most physical actions and interactions rather than in those involving just joysticks and controller buttons, specifically under a higher workload. 
Interestingly, eye parameters revealed to have the greatest influence on age detection in all actions but in the physical interactions, in which the controller position and rotation had higher impacts.

On gender profiling, instead, we observed how the walking activity was again the most prominent in helping detect the user's gender in AR, with the head position being the most influential sensor for detecting such personal information. 
Differently, in VR, our algorithms better performed during the pointing action and under actions at high workload. 
In this case, the eye-related behaviors demonstrated the most considerable influence on gender detection during both these actions. 
In agreement with AR findings, the head position is quite relevant. 
Both findings align with literature on the different eye and head movement behaviors between men and women.

\par
In conclusion, our work thoroughly studied users' profiling in AR and VR technologies. 
To the best of our knowledge, previous applied research on user profiling never compared performance obtained with these technologies.
On this matter, our results highlighted that profiling is more straightforward in Virtual Reality.
Through our ablation study, we additionally found eye sensors to be particularly useful in all our predictions (i.e., identification, age, gender), thus revealing to be likely responsible for the performance differences between AR and VR. 
Therefore, while being conscious of the technical challenges of accurately detecting eye behaviors in the real world, our findings highlight the importance of incorporating eye-tracking technologies to AR headsets.
To sum up, our work show the potential of user profiling methodologies with virtual technologies, and pave the road to several future works on how to improve AR and VR technologies with respect to users' profiling.

\bibliographystyle{unsrt}  
\bibliography{references}

\end{document}